\documentclass[twoside,11pt]{article}

 \usepackage{filecontents}

\usepackage{jmlr2e}

\usepackage[utf8]{inputenc}

\usepackage{amsmath}    
\usepackage{graphicx}   

\usepackage{times}
\usepackage{hyperref}
\usepackage{algorithm}
\usepackage{algpseudocode}
\usepackage{tikz,pgf}
\usepackage{epsfig,amssymb,amsmath,ifthen,comment,mathtools}
\usepackage{color}
\usepackage{array}
\usepackage{pdflscape}
\usepackage[figuresright]{rotating}
\usepackage{xr}

\usetikzlibrary{arrows,shapes.arrows,calc,shapes.geometric,shapes.multipart,  
decorations.pathmorphing,positioning,shapes.swigs}

\def\ci{{\perp\!\!\!\perp}}

\newcommand{\G}{{\mathcal G}}

\newcommand{\GVa}{{\cal G}\!(\!V\!(\!a\!)\!)}
\newcommand{\GYa}{{\cal G}\!(\!Y\!(\!a\!)\!)}

\newenvironment{prf}{\noindent\textit{Proof:}\begin{mdseries}}{\end{mdseries}{\hfill\scriptsize$\Box$}}

\newtheorem{thm}{Theorem}

\newtheorem{rmk}{Remark}
\newtheorem{exa}{Example}

\newtheorem{prop}{Proposition}

\DeclareMathOperator{\doo}{do}
\DeclareMathOperator{\dis}{dis}

\DeclareMathOperator{\pa}{pa}
\DeclareMathOperator{\de}{de}

\DeclareMathOperator{\an}{an}

\DeclareMathOperator{\mb}{mb}

\DeclareMathOperator{\pre}{pre}

\DeclareMathOperator{\pas}{pas}

\newcommand{\ilya}[1]{{\color{red!70!black} #1}}

\jmlrheading{1}{2021}{1-?}{8/15}{??/??}{robins21}{Ilya Shpitser, Thomas S. Richardson, and James M. Robins}

\begin{document}

\title{Multivariate Counterfactual Systems And Causal Graphical Models}

\author{\name Ilya Shpitser \email ilyas@cs.jhu.edu\\
       \addr Department of Computer Science\\
	Johns Hopkins University\\
       Baltimore, MD 21218, USA
	\AND
       \name Thomas S.\ Richardson \email thomasr@uw.edu\\
       \addr Department of Statistics\\
       University of Washington\\
	Seatle, WA 98195-4550, USA
       \AND
\name James M.\ Robins \email robins@hsph.harvard.edu\\
	\addr Harvard T. H. Chan School of Public Health\\
	Boston, MA 02115, USA
}

\editor{Anonymous}

\maketitle

\begin{abstract}
Among Judea Pearl's many contributions to causality and statistics, the graphical \hbox{{\em d}-separation} criterion,
and the {\em do}-calculus stand out. In this paper we show that \hbox{{\em d}-separation} provides direct insight into an earlier
causal model originally described in terms of potential outcomes and event trees. In turn, the resulting synthesis leads to a simplification of the {\em do}-calculus that clarifies and separates the underlying concepts, {and a simple counterfactual formulation of a complete identification algorithm in causal models with hidden variables.}
\end{abstract}





\section{Introduction}

For the last three decades Judea Pearl has been a leading advocate for the adoption of causal models throughout the sciences. {\citet{pearl95causal} introduced causal models based on  non-parametric structural equation models (NPSEMs).}%
\footnote{{See also \citep[p.69]{pearl09causality}. More recently Pearl has used the term (Structural) Causal Model (SCM) to refer to NPSEMs; see  \citep[p.203, Def.~7.1.1]{pearl09causality}. However, (S)CM is sometimes also used to denote NPSEMs in which, in addition, the error terms are assumed to be independent either explicitly, see
 \citep[p.44, Def.~2.2.2]{pearl09causality} \citep{forre19causal}, or implicitly see \citep{lee20generalized}. For this reason, we prefer to use Pearl's earlier terminology.
 }
}
{NPSEMs encode direct causal relations between variables. More precisely each variable $V$ is modeled as a function of its direct causes and an error term $\varepsilon_V$; this is the ``structural equation'' for $V$; see Table
\ref{fig:sem-table}. These causal relationships can be represented naturally by the directed arrows on a directed acyclic graph (DAG) in which there is an edge $X\rightarrow V$ if $X$ is present in the structural equation for $V$. The resulting graph is often called a causal DAG or diagram.
However, further probabilistic assumptions are required to link the
NPSEM to the distribution of the data.}

{Pearl has often considered a submodel of an NPSEM, hereafter referred to as the NPSEM-IE, which assumes the \ilya{i}ndependence of \ilya{e}rror terms.  NPSEM-IEs typically include both observed and hidden variables.%
\footnote{Causal DAG models with unobserved variables are also referred to as ``semi-Markovian'' by \citep[p.69 and p.76]{pearl09causality}.}
Thus although these models assume that errors are independent they still allow a modeler to postulate non-causal dependence between observed variables $X$ and $Y$ by including a hidden variable $X \leftarrow H \rightarrow Y$ (instead of allowing errors $\varepsilon_X$ and $\varepsilon_Y$ to be dependent).}

{Under the NPSEM-IE the distribution over the factual (i.e.~hidden plus observed) variables factorizes according to the causal DAG. This allows one to
reason about conditional independence in the distribution for the factual variables via d-separation relations on the
causal graph.  Based on this insight, Pearl developed an influential reasoning system called the {\it do}-calculus
which allows complex derivations to be made linking causal and observed quantities by appealing to d-separation in
graphs derived from the causal DAG.}

Causal graphs plus d-separation turn a difficult mathematical
problem into a simple one of graph topology.
The use of causal DAGs, as championed by Pearl, has revolutionized causal reasoning
in many fields, including fields such as epidemiology and sociology, precisely because causal
reasoning based on DAGs and d-separation is so ``user-friendly.'' That is,
individuals lacking the necessary mathematical background to understand
probabilistic inference based on solely on an NPSEM-IE have been given a tool
with which they can solve subtle problems in causal inference. In fact, even
the mathematically sophisticated find causal reasoning with graphs to be
much easier than algebraically manipulating the underlying structural
equations. As Pearl emphasizes this is largely because causal DAGs
faithfully represent the way humans, including scientists and mathematicians,
encode causal relations.

The use of DAGs to encode causal relationships dates back to
the work of the geneticist Sewall Wright \cite{wright21correlation} in the 1920s, who used a special case of the NPSEM associated with linear structural equations, and Gaussian errors for pedigree analysis among other applications in biology. 
These ideas were further developed and applied by Wright, Haavelmo, the Cowles Commission, Strotz \& Wold, Fisher 
\citep{wright21correlation,haavelmo43simultaneous,simon:1953,strotz:wold:recursive:1960,fisher:causation:1969,fish:corr:1970}.


In statistics, (non-graphical) causal inference models have a long history also dating back to the 1920s \citep{neyman23app,rubin74potential,robins86new}. These models are based on counterfactual variables (potential
outcomes) that encode the value the variable would have if, possibly contrary
to fact, a particular treatment had been given.  Causal contrasts in these models
compare the distributions of potential outcomes under two or more treatments. 

In general, these counterfactual models considered treatments or exposures at a single point in time.
Extending the framework introduced by Neyman to allow for treatment at multiple time-points, 
Robins introduced {\it causally interpretable structured tree graph} (CISTG) models.
These counterfactuals models, which were represented using event tree graphs, 
extended the point treatment model of \citet{neyman23app} to longitudinal studies with time-varying treatments, direct and indirect effects and feedback of one cause on another.

Pearl has noted that a NPSEM (even without assumptions on the distribution of the errors) implies the existence of potential outcomes and thus an NPSEM model also allows reasoning about counterfactuals; see \citep{pearlhalpern,pearlhalpern2}.
Indeed,
Robins and Richardson have shown that in fact a particular finest CISTG model (``as
detailed as the data'') is mathematically isomorphic to an NPSEM model in
the sense that any such CISTG model can be written as an (acyclic) NPSEM model and vice-versa.  
A finest CISTG ``as detailed as the data'' is a
counterfactual causal model in which all the underlying variables can be
intervened on -- an assumption that Pearl has sometimes also adopted.\footnote{{See \citep{galles98axiomatic}, Definitions~2 and 3 and footnote 2, also \citep{pearl09causality} Definitions~7.1.2 and 7.1.3. However,  in more recent work, 
\citet{pearl18does,pearl19on} has made a further distinction between hypothetical interventions and a concept of causation based on variables that ``listen to others.'' 
Pearl continues to assume that for every variable 
there are counterfactuals associated with applying the {\it do} operator to that variable. 
However, the model resulting from applying the {\it do} operator and removing structural equations need no longer correspond to an actual intervention.
This leaves open the question as to whether there are predictions made by these removals and, if so, how can they be validated. 
}
 }
Other versions of
CISTG models, unlike the NPSEM,  assume that only a subset of the
variables can be thought of as treatments with associated counterfactuals;
 thus interventions and causal effects are only defined for this subset. 
Henceforth, unless  stated otherwise, the term ``CISTG model'' will be used to denote a ``finest CISTG model as detailed as the data.''

Since counterfactual variables are not directly observed, assumptions are needed to link 
 counterfactuals and their distributions to those of the factual data.
A necessary assumption is consistency, which states that for a unit
their observed outcome ($Y$) and their potential outcome ($Y(a)$) had a particular treatment $a$ been assigned, will coincide, if in reality the treatment they received ($A$) is $a$.  
However, since both counterfactuals are not directly observed for any individual -- the fundamental problem of causal inference -- distributions of causal effects are not identified without additional assumptions, beyond consistency. 

These assumptions typically take the form of Markov 
(conditional independence) assumptions that link the distribution of the factual data to that of the counterfactuals, as further discussed below.  The simplest example is a randomized clinical trial which 
assigns treatment via the flip of a coin, and thus treatment is independent of the potential outcomes, so for all $a$, $A \ci Y(a)$. 
Together, consistency and Markov assumptions allow population level causal contrasts to be identified from observed data. 
In contrast, individual level effects are not typically identifiable.

Under the NPSEM-IE model, 
the additional Markov assumptions follow from the assumption that the errors in the structural  equation for each variable (hidden or observed)  are independent of the errors in the structural equations for the other variables.

\cite{robins86new} similarly added independence assumptions to the CISTG model. 
Robins referred to the version of this model in which all variables can be intervened on 
 as the ``finest fully randomized CISTG model as detailed as the data,'' which we henceforth refer to as the ``FFRCISTG model,'' unless stated otherwise. 
Interestingly, the NPSEM-IE implies many more counterfactual independence assumptions than does the
corresponding FFRCISTG model. In fact, if we consider complete graphs on $p$ binary variables then the
difference between the number of assumptions implied by the NPSEM-IE and the FFRCISTG model grows at a doubly exponential rate.\footnote{With three binary variables, the difference in the dimension of the two models is $94$,  with four it is $32,423$ \citep{thomas13swig}.\label{ftn:dim}}

The NPSEM-IE allows  
the identification of certain causal effects -- the pure and total direct and
indirect effects and more generally path specific effects\footnote{See \citep{rrs21volume_mediation_arxiv} for more detail on these effects.} -- by making use of additional independence assumptions that cannot be confirmed, even in principle, by any experiment conducted using the variables represented on the graph.
In contrast under the less restrictive  FFRCISTG model all counterfactual independence assumptions are in principle experimentally  testable,\footnote{This assumes that it is possible to observe the natural value of a variable and then intervene on it an instant later; see discussion in Section \ref{sec:po-npsem-g}
.} the pure and total direct effects are not identifiable (from the variables on the graph).
However, ordinary intervention distributions of the type that arise in Pearl's {\it do}-calculus are identifiable under the
FFRCISTG model.

Many statisticians and econometricians exclusively use counterfactuals (without graphs) when carrying out causal data analyses. 
Pearl has developed purely graphical criteria to reason about confounding and many other causal questions. 
Since graphical criteria, such as Pearl's {\it do}-calculus, make no reference to counterfactuals, they can appear confusing to those
unused to causal graphs. Indeed only factual variables typically appear on Pearl's causal
diagrams so any connection between Pearl's graphical criteria and the
statistician's counterfactual criteria appear at first glance to be obscure. This is true even
though Pearl and others have shown mathematically that the two approaches to
evaluation of confounding are effectively logically equivalent.

In this paper, we will describe an approach that unifies the graphical and
counterfactual approaches to causality, via a graph known as a Single-World Intervention Graph (SWIG).%
\footnote{\label{foot:twin}
The approach taken here is inspired by, but distinct from, earlier approaches to combining graphs and counterfactuals such as Pearl's Twin-Network approach  \citep{balke94countereval},
\citep{shpitser07hierarchy}, \citep[Section 7.1.4]{pearl09causality}.

However, the d-separation criterion on twin-networks is not complete, as there are 
deterministic relationships that are present -- but not represented graphically -- among the variables
in a twin-network.
Consequently, it is possible for there to be a d-connecting path and yet the corresponding conditional independence holds for all distributions in the model; see the Appendix for a simple example.
In contrast, d-separation is complete for a SWIG.
However, it should be noted that twin-networks are addressing a harder problem than SWIGs since their goal is to determine all independencies implied by an NPSEM-IE model including ``cross-world'' independencies. 

Finally, we note that twin-network graphs have not typically used (minimal) labelings, which turn out to be important in some applications; see Section \protect\ref{subsec:swig}.
}
The SWIG is defined by the counterfactual  independencies implied by the FFRCISTG model. 
The nodes on  a SWIG correspond to the counterfactual random variables present in these independences.
Furthermore, Pearl's d-separation criterion can be applied to the SWIG to read off counterfactual independences
implied by the FFRCISTG model.
 In fact, we will show that SWIGs lead directly to a simpler reformulation of the {\it do}-calculus 
 in terms of potential outcomes that allows a considerable simplification of Rule 3.
 This reformulated calculus, that we term the potential outcome calculus or {\it po}-calculus is also strictly stronger than Pearl's in that it may be used to infer equalities that are not expressible in terms of the {\it do}$(\cdot)$ operator. We use the {\it po}-calculus to derive a new simple formulation of an extended version of the ID algorithm for identification of causal queries in the presence of hidden variables.
 The extended algorithm identifies joint distributions over sets of counterfactual outcomes, where some outcomes are the ``natural'' values that treatment variables would take were they not intervened on. 





\section{Graphs, Non-Parametric Structural Equation Models (NPSEMs) And The {\it g-/do} Operator}
\label{sec:po-npsem-g}

Fix a set of indices $V \equiv \{ 1, \ldots, K\}$ under a total ordering $\prec$, define the sets $\pre_{i} \equiv \{ 1, \ldots, i-1 \}$.  For each index $i \in V$, associate a random variable $X_i$ with state space ${\mathfrak X}_i$; the ordering here could be given by temporal ordering but need not be.\footnote{If some variables do not affect variables later in time then many non-temporal orders may be used; see \cite[Chapter~11]{robins86new}  and later.}
 Given $A \subseteq V$, we will denote subsets of random variables indexed by $A$ as $X_A \in {\mathfrak X}_A \equiv \bigtimes_{i\in A}{\mathfrak X}_i$.
For notational conciseness we will sometimes use index sets to denote random variables themselves, using $V$ and $A$ to denote $X_V$ and $X_A$,  respectively, and similarly using lower case $a$ to denote $x_A \in {\mathfrak X}_A$. Similarly, by extension, we will also use $V_A$ to denote $X_A$ and $V_i$ to denote $X_i$.
 
We assume the existence of all one-step-ahead \emph{potential outcome} (also called counterfactual) random variables of the form $V_i(x_{\pa_{i}})$, where $\pa_{i}$ is a fixed subset of $\pre_{i}$, and $x_{\pa_{i}}$ is any element in ${\mathfrak X}_{\pa_{i}}$.\footnote{$\pa$ here is short for ``parent,'' which will be motivated later when we subsequently build a connection to directed graphs.}  The variable $V_i(x_{\pa_{i}})$ denotes the value of $V_i$ had the set $V_{\pa_{i}}$ of \emph{direct causes of $V_i$} been set, possibly contrary to fact, to values ${\pa_{i}}$.  The existence of a total ordering $\prec$ on indices and the fact that $\pa_{i} \subseteq \pre_{i}$ precludes the existence of cyclic causation.  That is, we consider causal models that are \emph{recursive}.
$V_i(x_{\pa_{i}})$ may be conceptualized as the output of a \emph{structural equation}
$f_i : (x_{\pa_{i}}, \epsilon_i ) \mapsto { x}_i$, a function representing a causal mechanism that maps values of $x_{\pa_{i}}$, as well as the value of a variable $\epsilon_i$, to values of $V_i$.
Specifically, we may define the error term $\epsilon_i$ to be the vector comprised of the set of random variables
 $\{V_i(x_{\pa_i}) \mid x_{\pa_i} \in {\mathfrak X}_{\pa_{i}} \}$
and  $f_i$ to be such that 
$f_{i}(x_{\pa_i},{\epsilon}_{i}) \equiv ({\epsilon}_{i})_{x_{{\pa}_i}} = 
V_i(x_{\pa_i})$.

We define {non-parametric structural equation} causal models (NPSEMs) as sets of densities over the set of random variables
\begin{align*}
\mathbb{V} \equiv \{ V_i(x_{\pa_{i}}) \mid i \in V,\; {x_{\pa_{i}}} \in {\mathfrak X}_{\pa_{i}} \}.
\end{align*}
Note that $\mathbb{V}$ includes variables $V_i$ which have no parents, and which are thus factual.
For simplicity of presentation, we assume ${\mathfrak X}_i$ is always finite, and thus ignore the measure-theoretic complications that arise with defining densities over sets of random variables in the case where some state spaces ${\mathfrak X}_{\pa_{i}}$ are infinite.

\begin{sidewaystable}

{\bf FFRCISTG/SWIG Potential Outcome Model \cite{robins86new,thomas13swig}}
\bigskip

\begin{center}
\begin{tabular}{llclclclclcl}
&\multicolumn{3}{l}{One-Step Ahead} &&   \multicolumn{3}{l}{Passive} && \multicolumn{3}{l}{Experimental}\\
&\multicolumn{3}{l}{Counterfactuals} &&   \multicolumn{3}{l}{Observation}&& \multicolumn{3}{l}{Intervention on $A$}\\
\\[-8pt]
{\bf Graph:}\\[-12pt]
&
\multicolumn{3}{p{3cm}}{
\begin{tikzpicture}[>=stealth, ultra thick, node distance=2cm,
    pre/.style={->,>=stealth,ultra thick,blue,line width = 1.2pt}]
     \tikzstyle{format} = [draw, ultra thick, minimum size=6mm, inner sep=1pt]
\begin{scope}
\node[format, thick, xshift=0cm, yshift=0.1cm, inner sep=1.2pt,
		          name=l, shape=swig vsplit, swig vsplit={gap=4pt, inner line width right=0.3pt}]{
        					\nodepart{left}{$L$}
        					\nodepart{right}{$l$}};
\node[format, thick, xshift=0.3cm, yshift=0.1cm, inner sep=1.2pt, below = 0.6 cm of l,
		          name=a, shape=swig vsplit, swig vsplit={gap=4pt, inner line width right=0.3pt}]{
        					\nodepart{left}{$A(l)$}
        					\nodepart{right}{$a$}};
\node[format, thick, inner sep=0.5pt, name=y,shape=ellipse,style={draw},right =0.4cm of a]
{$Y(a,l)$
};
\draw[pre,->] (l) to[out=300,in=80] (a);
\draw[pre,->] (l) to (y);
\draw[pre,->] (a) to (y);
\end{scope}
\end{tikzpicture}}
&&
\multicolumn{3}{p{3cm}}{
\begin{tikzpicture}[>=stealth, ultra thick, node distance=2cm,
    pre/.style={->,>=stealth,ultra thick,blue,line width = 1.2pt}]
     \tikzstyle{format} = [draw, ultra thick, minimum size=6mm, inner sep=1pt]
\begin{scope}
\node[format,thick,name=l,shape=circle,style={draw}] {
$L$
};
\node[format, thick, name=a,shape=circle,style={draw},below right=0.6cm and 0.1cm of l]
{$A$
};
\node[format, thick, name=y,shape=circle,style={draw},right =0.4cm of a]
{$Y$
};
\draw[pre,->] (l) to (a);
\draw[pre,->] (l) to (y);
\draw[pre,->] (a) to (y);
\end{scope}
\end{tikzpicture}}&&
\multicolumn{3}{p{3cm}}{
\begin{tikzpicture}[>=stealth, ultra thick, node distance=2cm,
    pre/.style={->,>=stealth,ultra thick,blue,line width = 1.2pt}]
     \tikzstyle{format} = [draw, ultra thick, minimum size=6mm, inner sep=1pt]
\begin{scope}
\node[format,thick,name=l,shape=circle,style={draw}] {
$L$
};
\node[format, thick, xshift=0cm, yshift=0.1cm, inner sep=1.2pt, below right=0.6cm and 0.1cm of l,
		          name=a, shape=swig vsplit, swig vsplit={gap=4pt, inner line width right=0.3pt}]{
        					\nodepart{left}{$A$}
        					\nodepart{right}{$a$}};
\node[format, thick, name=y,shape=ellipse,style={draw},right =0.4cm of a]
{$Y(a)$
};
\draw[pre,->] (l) to (a);
\draw[pre,->] (l) to (y);
\draw[pre,->] (a) to (y);
\end{scope}
\end{tikzpicture}}
\\[4pt]
{\bf Variables:} & $L$ & & &&  $L$ & & & & $L$ \\
& $A(l)$&&   &\resizebox{5mm}{!}{\textcolor{blue}{$\Rightarrow$}}\;\;\; &   ${A}$ &{$\equiv$} & {$A(L)$} & \resizebox{5mm}{!}{\textcolor{blue}{$\Rightarrow$}}\;\;\; & $A$ &$\equiv$ & $A(L)$ \\
&$Y(a,l)$& &  &&  $Y$ &$\equiv$ & $Y(A,L)$ & &$Y(a)$ & $\equiv$ & $Y(a,L)$\\[6pt]
{\bf Interpretation:} &\multicolumn{3}{p{3cm}}{Counterfactuals when $A$ and $L$ are intervened on;} &&\multicolumn{3}{l}{Observed System;}&&
\multicolumn{3}{p{3cm}}{ Prior to intervention: $L$ and $A$; after intervention: $Y(a)$.}
\\[30pt]
{\bf Meaning of} $A${\bf :} &\multicolumn{3}{m{3cm}}{($A$ does not appear)}&&\multicolumn{3}{l}{Natural value of $A$;}&& \multicolumn{3}{p{3cm}}{Natural value of $A$ (observed prior to intervention on $A$).}\\
\end{tabular}\\[15pt]
\begin{tabular}{ll}
{\bf Single World No Confounding Assumption:}& for each pair $a,l$:\quad $L \;\ci\; A(l)  \;\ci\; Y(a,l)$\\[20pt]
\end{tabular}
\end{center}
\caption{Graphical causal models based on the SWIG/FFRCISTG counterfactual framework. \label{fig:swig-table}
 Note that the meaning of a variable, such as $A$  or $Y$ does not depend on the graph in which it appears; compare to Table \protect\ref{fig:sem-table}.
 The SWIG no confounding assumption is less restrictive than the assumption of independent errors.}
\end{sidewaystable}

\begin{sidewaystable}
{\bf Non-Parametric Structural Equation Model with Independent Errors (NPSEM-IE)  \cite{strotz:wold:recursive:1960,pearl09causality}}
\bigskip

\begin{center}
\begin{tabular}{llclclclclcl}
&\multicolumn{3}{l}{Passive} &&   \multicolumn{3}{l}{Experimental} &&   \multicolumn{3}{l}{Experimental}\\
&\multicolumn{3}{l}{Observation} &&   \multicolumn{3}{l}{Intervention on $A$} &&   \multicolumn{3}{l}{Intervention on $A$ and $L$}\\
\\[2pt]
{\bf Graph:}\\[-20pt]
&
\multicolumn{3}{p{3cm}}{
\begin{tikzpicture}[>=stealth, ultra thick, node distance=2cm,
    pre/.style={->,>=stealth,ultra thick,blue,line width = 1.2pt}]
     \tikzstyle{format} = [draw, ultra thick, minimum size=6mm, inner sep=1pt]
\begin{scope}
\node[format,thick,name=l,shape=circle,style={draw}] {
$L$
};
\node[format, thick, name=a,shape=circle,style={draw},below right=0.6cm and 0.1cm of l]
{$A$
};
\node[format, thick, name=y,shape=circle,style={draw},right =0.4cm of a]
{$Y$
};
\draw[pre,->] (l) to (a);
\draw[pre,->] (l) to (y);
\draw[pre,->] (a) to (y);
\end{scope}
\end{tikzpicture}}
&&
\multicolumn{3}{p{3cm}}{
\begin{tikzpicture}[>=stealth, ultra thick, node distance=2cm,
    pre/.style={->,>=stealth,ultra thick,blue,line width = 1.2pt}]
     \tikzstyle{format} = [draw, ultra thick, minimum size=6mm, inner sep=1pt]
\begin{scope}
\node[format,thick,name=l,shape=circle,style={draw}] {
$L$
};
\node[format, thick, name=a,inner sep=0pt, shape=rectangle,style={draw},below right=0.6cm and 0.1cm of l, red]
{$a$
};
\node[format, thick, name=y,shape=circle,style={draw},right =0.4cm of a]
{$Y$
};
\draw[pre,->] (l) to (y);
\draw[pre,->] (a) to (y);
\end{scope}
\end{tikzpicture}}&&
\multicolumn{3}{p{3cm}}{
\begin{tikzpicture}[>=stealth, ultra thick, node distance=2cm,
    pre/.style={->,>=stealth,ultra thick,blue,line width = 1.2pt}]
     \tikzstyle{format} = [draw, ultra thick, minimum size=6mm, inner sep=1pt]
\begin{scope}
\node[format,thick,name=l,shape=rectangle,style={draw}, red] {
$l$
};
\node[format, thick, name=a,shape=rectangle,style={draw},below right=0.6cm and 0.1cm of l, red]
{$a$
};
\node[format, thick, name=y,shape=ellipse,style={draw},right =0.4cm of a]
{$Y$
};
\draw[pre,->] (l) to (y);
\draw[pre,->] (a) to (y);
\end{scope}
\end{tikzpicture}}
\\[4pt]
{\bf Variables:} &$L$ &$=$& $f_L(\varepsilon_L)$ &&  $L$ &$=$ & $f_L(\varepsilon_L)$ & & \textcolor{red}{$L$} &\textcolor{red}{$=$} & \textcolor{red}{$l$} \\
&$A$ &$=$&  $f_A(L, \varepsilon_A)$ & \resizebox{5mm}{!}{\textcolor{blue}{$\Rightarrow$}}\;\;\; &   $\textcolor{red}{A}$ &\textcolor{red}{$=$} & \textcolor{red}{$a$} 
& \resizebox{5mm}{!}{\textcolor{blue}{$\Rightarrow$}}\;\;\; &   $\textcolor{red}{A}$ &\textcolor{red}{$=$} & \textcolor{red}{$a$}\\
&$Y$ &$=$ & $f_Y(A,L,\varepsilon_Y)$ &&  $Y$ &$=$ & $f_Y(A,L,\varepsilon_Y)$&& $Y$ &$=$ & $f_Y(A,L,\varepsilon_Y)$\\[6pt]
{\bf Interpretation:} &\multicolumn{3}{l}{Observed system;} &&\multicolumn{3}{p{3cm}}{Variables in system in which $A$ is set to $a$;}&&
\multicolumn{3}{p{3cm}}{\small $L$ and $A$ after each is intervened on; $Y$ after both interventions.}
\\[28pt]
{\bf Meaning of} $A${\bf :} &\multicolumn{3}{l}{Natural value of $A$;}&&\multicolumn{3}{p{3cm}}{Value of $A$ after intervention on $A$;}
&&\multicolumn{3}{p{3cm}}{{\small Value of $A$ after intervention on $A$ (and $L$).}}
\end{tabular}\\[20pt]
\begin{tabular}{lllll}
\multicolumn{4}{l}{\bf Independent Errors No Confounding Assumption:\quad $\varepsilon_L \;\ci\; \varepsilon_A  \;\ci\; \varepsilon_Y$} \\[12pt]
\multicolumn{2}{l}{\bf Relationships to Counterfactuals:}&\\[6pt]
\kern30pt &Error terms: & $\varepsilon_L = L$; &$\varepsilon_A = \{A(l) \hbox{ for all } l \}$; &$\varepsilon_Y = \{Y(a,l) \hbox{ for all } l,a \}$\\[8pt]
&Structural equations: & $L=f_L(\varepsilon_L)$  & $A(l) = f_A(l,\varepsilon_A)$ & $Y(a,l) = f_Y(a,l,\varepsilon_Y)$
\end{tabular}
\bigskip

\end{center}
\caption{Structural equation models and their relationship to counterfactuals.
 Note that the meaning of a variable such as $A$ or $Y$ is context specific, it depends on the graph in which it appears.
\label{fig:sem-table}
This NPSEM is a strict sub-model of the SWIG given in Table \protect\ref{fig:swig-table}. This is (solely) because the SWIG no confounding assumption is less restrictive than the assumption of independent errors.}
\end{sidewaystable}

Given a set of one-step-ahead potential outcomes $\mathbb{V}$, for any $A \subseteq V$ and $i \in V$,
the potential outcome $V_i(a)$, the response of $V_i$ had variables in $V_A$ been set to $a \in {\mathfrak X}_A$, is the one step ahead counterfactual $V_i(\pa_i) \in \mathbb{V}$ if $V_A = V_{\pa_i}$, and is otherwise defined via \emph{recursive substitution}:
\begin{align}
V_i(a) \equiv V_i\!\left(a_{\pa_i},\, V_{ \pa_i\! \setminus A}(a)\right).
\label{eqn:rec-sub}
\end{align}
In words, this states that $V_i(a)$ is the potential outcome where variables in both ${\pa_i}$ and $A$ are set to their corresponding values in $a$, and all elements of $\pa_i$ not in $A$ are set to whatever values their recursively defined counterfactual versions would have had had $V_A$ been set to $a$.  This is well-defined because of of the requirement that 
$\pa_i\subseteq \pre_i$.

Equivalently, $V_i(a)$ is the random variable induced by a modified set of structural equations: specifically the set of functions $f_j$ for $V_j$
such that $A \cap \pa_j \neq \emptyset$ are replaced by modified functions 
$f_j^a : (x_{\pa_j \setminus A}, \varepsilon_j) \mapsto x_j$
that are obtained from 
$f_j : (x_{\pa_j}, \varepsilon_j) \mapsto x_j$
by always evaluating $\pa_j \cap A$ at the corresponding values in $a$.

We will extend our notational shorthand by using index sets to denote sets of potential outcomes themselves.  Thus, for $B\subset V$, we let $B(a)$ denote the set of potential outcomes $V_B(a)$.  
We denote by ${\mathbb V}^*$ the set of all variables derived by (\ref{eqn:rec-sub}) from ${\mathbb V}$ for all possible choices of the set $A$ (together with the set ${\mathbb V}$ itself).%
\footnote{The set ${\mathbb V}^*$ corresponds to Robins' {\it Finest Causally Interpreted Structured Tree Graph as Detailed as the Data}.
See Appendix C in \citep{thomas13swig}.}


While the potential outcome and the structural equation formalisms both yield the same causal model, there are some differences regarding the way in which the frameworks are typically presented. Specifically, regarding which ``objects'' are taken as primitive and which are derived.

The counterfactual formalism here starts with one-step-ahead counterfactuals that intervene on every parent (direct cause) of every variable, and constructs all other counterfactuals by means of recursive substitution.  Recursive substitution implies, in particular,  that $A(a)\equiv A$. This accords with the substantive claim  that it is possible to {\it first}  learn the ``natural'' value a variable $A$ would take on, and {\it then} an instant {\it later} intervene setting it to a specific value $a$, resulting in all subsequent variables $V_i$ behaving as counterfactual variables $V_i(a)$.

On the other hand, the structural equation formalism typically starts with a set of unaltered structural equations which yields the observed data distribution (via substitution).  Counterfactual distributions representing an intervention that sets elements in $A$ to $a$ are generated by replacing structural equations corresponding to elements in $A$ by degenerate functions that yield constants in $a$ \citep{strotz:wold:recursive:1960,pearl09causality}.  The resulting modified equation system thus represents the set of variables (including $A$)  {\it after} the action of setting $A$ to the value $a$. 
Consequently, there are two subtle, but important notational (not conceptual) distinctions:
\begin{itemize}
\item Under the standard presentation of structural equation models, used by Pearl, the meaning of a variable such as $A$, $Y$ or $L$, is dependent on the set of equations in which it appears. For example in Table \ref{fig:sem-table}, $Y$ in the left (unmodified) display corresponds to the natural value; in the middle display $Y$ corresponds to the value after intervening on $a$ or $Y(a)$ in counterfactual notation; in the right display $Y$ denotes the value after intervening on $A$ and $L$, or $Y(a,l)$.
In contrast, in the potential outcome framework, variables that are affected by an intervention take on a new name. 
\item Second, in the standard presentation, the variable ``$A$'' in the modified set of equations represents the variable after it has been intervened on. Thus for Pearl, $\hbox{do}(A=a)$ implies that $A=a$, a property he terms ``effectiveness.''\footnote{If we were to use $A(a)$ to designate the value taken by a variable $A$ {\it after} an intervention on $A$, then we could express this as $A(a)=a$. However, as noted, we use $A(a)$ to designate the value taken by a variable $A$ immediately prior to the intervention.}   
\end{itemize}

In this paper we will follow the notation conventions that are used in the potential outcome framework, but we stress that formally, non-parametric structural equations (NPSEMs) and one-step ahead counterfactuals are equivalent conceptually.\footnote{However, as described below, structural equation models are often used under an additional (strong) assumption of independent errors. Since this is a stronger assumption than typically used in the potential outcome framework, we use the acronyms NPSEM and NPSEM-IE to distinguish whether this additional assumption is being made.} See the {\it Variables} rows in Tables~\ref{fig:swig-table} and \ref{fig:sem-table} to see the correspondence between sets of one-step ahead counterfactuals and systems of structural equations; see \citep{pearl95causal,imbens2014} for further discussion of the representation of structural equations via potential outcomes.

Given a set $A \subseteq V$, the distribution on $V \setminus A$ resulting from setting $A$ to $a$ by interventions has been denoted in \citep{robins86new} by $p(V \setminus A \,|\, g=a)$, and subsequently as $p(V \setminus A \,|\, \doo(a))$ \citep{pearl09causality}.  The potential outcome view also allows us to consider distributions $p(V(a))$ for any $A \subseteq V$.  In such a distribution, variables in $A$ occur both as random and intervened on versions.  We later consider identification theory for distributions of this sort, where the set of treatment variables and outcome variables may intersect.

Recursive substitution in NPSEMs provides a link between observed variables and potential outcomes. In particular, it implies the \emph{consistency property}: for any disjoint $A,B \subseteq V$, $i \in V \setminus (A \cup B)$, $a \in {\mathfrak X}_A$, $b \in {\mathfrak X}_B$,
\begin{align}
V_B(a) = b\text{ implies } V_i(a,b) = V_i(a).
\label{eqn:consist}
\end{align}
See \citep{robins86new,robins87newerrata} and, for a proof using notation similar to this paper, \citep{thomas13swig,malinsky19po}.  Consistency is often phrased in a simpler form where $A = \emptyset$, yielding the identity $V_i(b) = V_i$ if $V_B = b$.

Equation (\ref{eqn:rec-sub}) also implies the \emph{causal irrelevance property}, namely that every $V_i(a)$ can be written as a function of a unique minimally causally relevant subset of $a$, as follows. (See  \cite{robins86new,thomas13swig} and, for a formulation similar to that used here, \cite{malinsky19po}.) Given ${\mathbb V}^*$ derived from ${\mathbb V}$ via (\ref{eqn:rec-sub}), let $V_i(a) \in \mathbb{V}^*$,
and let $A^*$ be the maximal subset of $A$ such that for every $j \in A^*$,
there exists a sequence ${w_1}, \ldots, {w_m}$ that does not intersect
$A$, where $j \in \pa_{w_1}$, ${w_i} \in \pa_{w_{i+1}}$, for $i = 1, \ldots m-1$, and ${w_m} \in \pa_i$.
Then, $V_i(a) = V_i(a^*)$.



As an example, given the indices $\{ 1, 2, 3 \}$, under the ordering $1 \prec 2 \prec 3$, if $\pa_2 = \{ 1 \}$, and $\pa_3 = \{ 2 \}$, we have one step ahead potential outcomes $V_1, V_2(v_1), V_3(v_2)$, for any values $v_1,v_2$.  We can define other counterfactuals via (\ref{eqn:rec-sub}), for example $V_3(v_1) \equiv V_3(V_2(v_1))$.  Consistency implies statements of the form $V_1 = v_1 \Rightarrow V_2 = V_2(v_1)$, while causal irrelevance implies $V_3(v_2, v_1) = V_3(v_2)$.

Both consistency and causal irrelevance hold in any NPSEM, in the sense that these properties are implied by the existence of a total order on variables we wish to consider, the existence of one step ahead counterfactuals, and (\ref{eqn:rec-sub}).
While useful, these properties on their own fail to capture many of the hypotheses that arise in causal inference problems (either by design or assumption). These additional constraints correspond to conditional independence restrictions concerning the error terms in non-parametric structural equations.  Although causal models are well-defined without reference to graphs, much conceptual clarity may be gained by viewing them graphically.
Thus, before describing causal models in detail, we introduce graphs and graphical models.


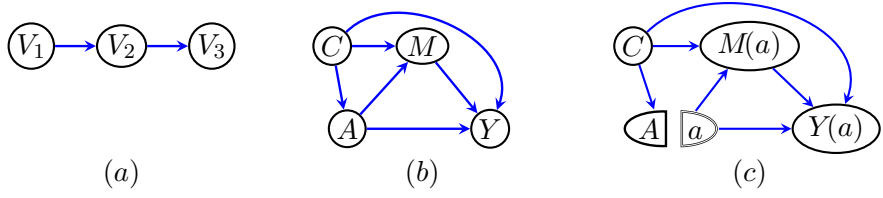
\begin{figure*}
	\begin{center}
		\begin{tikzpicture}[>=stealth, node distance=1.2cm]
		\tikzstyle{format} = [draw, thick, circle, minimum size=4.0mm,
		inner sep=1pt]
		\tikzstyle{unode} = [draw, thick, circle, minimum size=1.0mm,
		inner sep=0pt,outer sep=0.9pt]
		\tikzstyle{square} = [draw, very thick, rectangle, minimum size=4mm]

		\begin{scope}
		\path[->,  line width=0.9pt]
		node[format] (1) {$V_1$}
		node[format, shape=ellipse, right of=1, xshift=0.0cm] (2) {$V_2$}			
		node[format, right of=2, xshift=0cm,yshift=0cm] (3) {$V_3$}

		(1) edge[blue] (2)
		(2) edge[blue] (3)		
		
		node[below of=2, yshift=-0.5cm, xshift=0.0cm] (l) {$(a)$}
		;
		\end{scope}

		\begin{scope}[xshift=4.0cm]
		\path[->,  line width=0.9pt]
		node[format] (c) {$C$}
		node[format, shape=ellipse, right of=c, xshift=0.0cm] (m) {$M$}			
		node[format, below of=c, xshift=0.2cm,yshift=0.1cm] (a) {$A$}
		node[format, xshift=0.7cm, right of=a] (y) {$Y$}
		
		(c) edge[blue] (a)
		(a) edge[blue] (y)
		(a) edge[blue] (m)
		(m) edge[blue] (y)
		(c) edge[blue] (m)
		(c) edge[blue, out=45, in=70, looseness=1.1] (y)
		
		node[below of=m, yshift=-0.5cm, xshift=0.0cm] (l) {$(b)$}
		;
		\end{scope}
		
		\begin{scope}[xshift=8.0cm]
		\path[->, line width=0.9pt]
		node[format] (c) {$C$}
		node[format, shape=ellipse,right of=c, xshift=0.35cm] (m) {$M\!(a)$};
		\begin{scope}
			\tikzset{line width=0.9pt, inner sep=1.8pt, swig vsplit={gap=6pt, inner line width right=0.3pt}}	
				\node[below of=c, xshift=0.5cm, yshift=0.1cm, name=a, shape=swig vsplit]{
        					\nodepart{left}{$A$}
        					\nodepart{right}{$a$} };
		\end{scope}
		\path[->, thick]
		node[format, shape=ellipse, right of=a, xshift=1.0cm] (y) {$Y\!(a)$}
		
		(c) edge[blue] (a)
		(a) edge[blue] (y)
		(a) edge[blue] (m)
		(m) edge[blue] (y)
		(c) edge[blue] (m)
		(c) edge[blue, out=45, in=70, looseness=1.1] (y)
		
		node[below of=m, yshift=-0.5cm, xshift=0.0cm] (l) {$(c)$}
		;
		\end{scope}
		
		\end{tikzpicture}
	\end{center}
	\caption{(a) A DAG representing a simple NPSEM.
		(b) A simple causal DAG $\cal G$, with a treatment $A$, an outcome $Y$, a vector $C$ of baseline variables, and a mediator $M$.
		(c) A SWIG ${\cal G}(a)$ derived from (a) corresponding to the world where $A$ is intervened on to value $a$.
	}
	\label{fig:triangle}
\end{figure*}

\subsection{Graphical Models}
\label{subsec:graphs}

Statistical and causal models can be associated with graphs, where vertices represent variables, and edges represent (potential) statistical or causal relationships.  Formally, random variables are indexed by vertices. However, when we depict graphs we will display them with the random variables as vertices.

We will consider graphs with either directed edges only ($\to$), or mixed graphs with both directed and bidirected ($\leftrightarrow$) edges. 
Bidirected edges naturally arise as a way to represent (classes of) DAGs with latent variables; see Section \ref{subsec:latent} below.
In all cases we will require the absence of directed cycles, meaning that whenever the graph contains a path of the form $V_i \to \cdots \to V_j$, the edge
$V_j \to V_i$ cannot exist.  Directed graphs with this property are called directed acyclic graphs (DAGs), and mixed graphs with this property are called Acyclic Directed Mixed Graphs (ADMGs).  We will refer to graphs by ${\cal G}(V)$, where $V$ is the set of random variables indexed by $\{ 1, \ldots, K \}$.  We will write ${\cal G}$ in place of ${\cal G}(V)$ when the vertex set is clear.
We will use the following standard definitions for sets of vertices in a graph:

\begin{align}
\notag
\pa_i^{\cal G} \equiv& \{ j \,|\, V_j \to V_i \text{ in }{\cal G} \} & \text{ (parents of $V_i$)}\\
\notag
\an_i^{\cal G} \equiv& \{ j \,|\, V_j \to \cdots \to V_i \text{ in }{\cal G}, \hbox{or } V_j=V_i \} & \text{ (ancestors of $V_i$)}\\
\notag
\de_i^{\cal G} \equiv& \{ j \,|\, V_j \gets \cdots \gets V_i \text{ in }{\cal G}, \hbox{or } V_j=V_i \} & \text{ (descendants of $V_i$)}\\
\notag
\dis^{\cal G}_i \equiv& \{ j \,|\, V_j \leftrightarrow \cdots \leftrightarrow V_i \text{ in }{\cal G}, \hbox{or } V_j=V_i \} & \text{ (the district of $V_i$)}\\
\notag
\mb^{\cal G}_i \equiv& \{ j \,|\, V_j \leftrightarrow \cdots \leftrightarrow V_i \text{ in }{\cal G} \} \cup&\\
&	\{ j \,|\, V_j \to \circ \leftrightarrow \cdots \leftrightarrow V_i \text{ in }{\cal G} \}& \text{ (the Markov blanket of $V_i$).\footnotemark
}
\label{eqn:vertex-rels}
\end{align}
\footnotetext{Other authors often define the Markov blanket for a variable to be the minimal set $M$ that makes $V_i$ m-separated from $V\setminus (\{V_i\} \cup M)$.
Our definition corresponds to the minimal set $M$ such that $V_i$ is m-separated from its non-descendants.}
We will generally drop the superscript ${\cal G}$ if the relevant graph is obvious.
By definition, $\an_i^{\cal G} \cap \de_i^{\cal G} \cap \dis_i^{\cal G} = \{ V_i \}$.  We define these relations on sets disjunctively.
For example $\an_A^{\cal G} \equiv \bigcup_{V_i \in A} \an_i^{\cal G}$.

Given a DAG ${\cal G}(V)$, a statistical DAG model (also called a Bayesian network) associated with ${\cal G}(V)$ is a set of distributions that factorize (equivalently are Markov) with respect to ${\cal G}(V)$:
\begin{align}
p(V) = \prod_{i = 1}^K p(V_i \,|\, V_{\pa^{\cal G}_i}).
\label{eqn:d}
\end{align}

Given a distribution $p(V)$ that factorizes relative to a DAG $\mathcal{G}(V)$, conditional independence relations that are implied in $p(V)$ by (\ref{eqn:d}) can be derived using the well-known d-separation criterion \cite{pearl88probabilistic}. 
More precisely, if $p(V)$ is Markov relative to ${\cal G}(V)$, then the following \emph{global Markov property} holds: for any disjoint $X,Y,Z$ (where $Z$ may be empty)
\[
(X \ci_d Y \,|\, Z)_{{\cal G}(V)}\;\; \Rightarrow\;\; (X \ci Y \,|\, Z)_{p(V)}.
\]
Here $(X \ci_d Y \,|\, Z)_{{\cal G}(V)}$ denotes that $X$ is d-separated from $Y$ given $Z$ in ${\cal G}(V)$;
$ (X \ci Y \,|\, Z)_{p(V)}$ indicates that $X$ is independent of $Y$ given $Z$ in  $p(V)$.

The global Markov property given by d-separation allows reasoning about conditional independence restrictions implied by the statistical DAG model using qualitative, visual reasoning on paths in the graph.  

\subsection{Causal Models Associated With DAGs}

NPSEMs may be associated with directed graphs as well, by associating vertices with indices, and edges with relations given by $\pa_i, i \in \{ 1, \ldots, k \}$.  Specifically, given a (recursive) non-parametric structural equation model defined on $\mathbb{V}$ given the sets $\{ \pa_i | i \in \{ 1, \ldots, k \} \}$, we construct a \emph{causal diagram}, a directed acyclic graph (DAG) ${\cal G}(V)$ with a vertex for every $V_i$, $i \in \{ 1, \ldots, k \}$, and a directed edge from $V_j$ to $V_i$ if $j \in \pa_i$.  In other words, ${\cal G}(V)$ is defined by the NPSEM by letting $\pa_i^{\cal G} \equiv \pa_i$ for every $i$.  As an example, the NPSEM defined on the indices $\{ 1, 2, 3 \}$ described in the previous section corresponds to the DAG in Figure \ref{fig:triangle} (a).
 See the {\it Graph} rows of Tables~\ref{fig:swig-table} and \ref{fig:sem-table} for graphs corresponding to one-step ahead counterfactuals and structural equations.

Substantive knowledge may motivate additional independence assumptions relating to the set of one-step ahead counterfactuals $\mathbb{V}$.
As we will show below, such assumptions may also allow causal effects to be identified even when hidden variables are present.
Below we introduce two sets of such assumptions


\subsubsection*{Non-Parametric Structural Equations With Independent Errors}

A \emph{non-parametric structural equation model with independent errors}, or NPSEM-IE, is the set of distributions such that the $K$ different sets of one-step ahead variables satisfy:
\begin{align}
\{ V_1 \} \;\ci\; \left\{ V_2(x_{\pa_2}) \,|\, x_{\pa_2} \in {\mathfrak X}_{\pa_2} \right\}  \ci \cdots \ci \left\{ V_K(x_{\pa_K}) \,|\, x_{\pa_K} \in {\mathfrak X}_{\pa_K} \right\}
\label{eqn:mwm}
\end{align}
so that they are mutually independent of one another.   Phrased in terms of structural equations 
$f_i : (x_{\pa_j}, \varepsilon_i) \mapsto x_i$
for each $V_i$, the NPSEM-IE states that the joint distribution of the disturbance terms factorizes into a product of marginals:
$p(\epsilon_1, \ldots, \epsilon_K) = \prod_{i=1}^K p(\epsilon_i).$

NPSEMs with independent errors arise naturally as putative data-generating processes for a closed system.
For example, if we are simulating every variable in a model  then, it is natural to do this in a step-wise process 
 by specifying a set of structural equations. The equations provide recipes for generating a value for each variable in turn, 
given the previous values that have already been simulated plus an independently simulated error term.\footnote{Note that 
an NPSEM-IE may also contain unobserved variables, so that they include models described by Pearl as semi-Markovian \citep{pearl09causality}.}
See Table \ref{fig:sem-table} for an example. 

However, from an empiricist point of view the assumption of independent errors may be regarded as stronger than necessary. In particular, this assumption permits the identification of causal contrasts that are not subject to experimental verification even in principle;\footnote{Specifically, even if it were possible to carry out a randomized experiment manipulating any subset of the variables in the system, we could not directly observe certain counterfactual contrasts that are identified via an NPSEM-IE.} 
see the discussion in the companion paper \citep[Section \protect\ref{sec:mediation}]{rrs21volume_mediation_arxiv}.
 At the same time, many causal contrasts of interest, including all intervention distributions, may be identified under a much smaller set of assumptions.



 \subsubsection*{A Less Restrictive Model: Non-Parametric Structural Equations with Single World 
 (FFRCISTG) Independences}
 
 The above observations motivate an alternative approach based on
 the \emph{finest fully randomized causally interpretable structured tree graph (as detailed as the data)}, or, FFRCISTG model of Robins \cite{robins86new}.
  
The FFRCISTG model is ontologically liberal, but epistemologically conservative. Specifically, all the counterfactual queries that may be formulated within the scope of a non-parametric structural equation model, are still well-defined under this alternative, but, in contrast to the NPSEM-IE, only  those contrasts that could in principle be experimentally verified by experiments on the variables in the system are identified.


%
%

A non-parametric structural equation model with FFRCISTG independences is the set of counterfactual distributions satisfying 
\begin{align}
\hbox{\it For each }x_V \in {\mathfrak X}_V, \hbox{\it we have }
V_1\;\ci \; V_2(x_{\pa_2}) \;\ci\; \cdots \;\ci\;V_K(x_{\pa_K});
\label{eqn:swm}
\end{align}
see \citep{robins10alternative}. Thus for each $x_V \in {\mathfrak X}_V$ there is a set of $K$ random variables (the $K$ one-step ahead counterfactuals associated with $x_V$) and the variables {\it within} each such set are assumed to be mutually independent.  As $V_1$ is first in the ordering, it has no parents.


The FFRCISTG assumptions could be empirically verified in a set of randomized experiments, one for each $x_V$, under which we intervene on every variable in turn,  setting $V_i$ to the value $x_i$, but 
just before doing so, we are able to observe the random variable $V_i(x_{\pa_i})$, resulting from our earlier interventions. (Here it is assumed that because we intervene to set $V_i$ to $x_i$ an instant after it is measured, the value $V_i(x_{\pa_i})$ does not causally influence any subsequent variable.) Note the counterfactual random variables  in (\ref{eqn:swm}) all refer to a specific set of values $x_V$, which thus correspond to a single counterfactual ``world''.

Note that (\ref{eqn:mwm}) imposes all restrictions in (\ref{eqn:swm}), and in general exponentially many more.\footnote{In fact, in  the case where all variables are binary, the fraction of experimentally untestable constraints implied by the NPSEM-IE rises at a doubly exponential rate in the number of variables. See \citep[Section 5]{thomas13swig} and Footnote \ref{ftn:dim}.} Thus the FFRCISTG is less restrictive than the NPSEM-IE model, in other words, the NPSEM-IE is a strict submodel of the FFRCISTG.


As an example, the NPSEM associated with Figure \ref{fig:triangle} (b) is defined using one step ahead counterfactuals $C$, $A(c)$, $M(c,a)$, and
$Y(c,a,m)$, for every value set $c,a,m$.  Then the FFRCISTG model restrictions for this NPSEM imply that 
\begin{align}
\hbox{\it For each set of values } c,a,m,\;\;C \ci A(c) \ci M(a,c) \ci Y(c,a,m),
\label{eqn:swm2}
\end{align}
while the NPSEM-IE restrictions for the NPSEM state that
\begin{align}
\hbox{\it For each set of values }c,c',c'',a,a',\;\; C \ci A(c) \ci M(a,c') \ci Y(c'',a',m).
\label{eqn:mwm2}
\end{align}
The restrictions in Equation (\ref{eqn:swm2}) are a strict subset of the restrictions in Equation (\ref{eqn:mwm2}) which are themselves a subset of the restrictions defining the NPSEM-IE.

Interventional distributions of the form $p(V(a))$, for $A \subseteq V$ in both of the above models may be represented in graphical form by 
a simple splitting operation on DAGs.  The graphs resulting from this operation will be called
Single World Intervention Graphs (SWIGs) \citep{thomas13swig} for reasons that will be described below.

\subsection{Single World Intervention Graphs (SWIGs)}
\label{subsec:swig}

SWIGs were introduced in \citep{thomas13swig} as graphical representations of potential outcome distributions that help unify the graphical and potential outcome formalisms.
Given a set $A$ of variables and an assignment $a$ to those variables, a SWIG ${\cal G}(V(a))$ may be constructed
from ${\cal G}(V)$ by splitting all vertices in $A$ into a random half and a fixed half, with the random half inheriting all  edges with an incoming arrowhead and the fixed half inheriting all outgoing directed edges.  Then, all random vertices $V_i$ are re-labelled as $V_i(a)$ or equivalently (due to causal irrelevance) as $V_i(a_{\an^*_i})$, where $\an^{*}_i$ consists of the fixed vertices that are ancestors of $V_i$ in the split graph; the latter labelling is referred to as the {\it minimal labelling} of the SWIG.
By using minimal labelling the SWIG encodes the property of causal irrelevance, so that, for example, if $Y(x)$ appears in ${\cal G}(x,z)$ then 
$Y(x,z)= Y(x)$.

Fixed nodes are enclosed by a double line.  For an example of a SWIG representing the joint density $p(Y(a), M(a), C(a), A(a)) = p(Y(a), M(a),C,A)$, 
under the FFRCISTG model (and thus under an NPSEM-IE) associated with the DAG of Figure \ref{fig:triangle} (b) see Figure \ref{fig:triangle} (c).  
If the vertex set $V$ is assumed or obvious, we will denote ${\cal G}(V(a))$ by ${\cal G}(a)$,
just as ${\cal G}(V)$ is denoted by ${\cal G}$.

Thus a SWIG ${\cal G}(V(a))$ is a DAG with vertex set $V(a) \cup a$; 
the vertices in $V(a)$ correspond to random variables, while vertices in $a$ are fixed, taking a specific value.

In a SWIG, every treatment variable has two versions: a fixed version representing the intervention on that treatment, and a random version (which corresponds to measuring the treatment variable just before the intervention took place).  This feature of SWIGs allows them to directly express, using d-separation, independence restrictions linking observed versions of treatments, and counterfactual variables representing responses after treatments have been set.

Restrictions of this type, which generalize the well known conditional ignorability restriction\footnote{Specifically, $Y(a) \ci A \mid C$, for some set of baseline covariates $C$.} will be used later to reformulate the second rule of the do-calculus, using the language of SWIGs and potential outcomes.

Pearl's ``mutilated graphs,'' which are an alternative graphical representation of interventional distributions, only contain the fixed versions of treatments.  This makes it difficult to express restrictions such as conditional ignorability.  Instead, the do-calculus uses a variant of the mutilated graph where certain outgoing edges are also removed.  An additional difficulty with this variant, though it is formally correct, is that vertices on it are not labelled as counterfactual random variables.

Tables \ref{fig:sem-table} and \ref{fig:swig-table} illustrate, via simple examples, how SWIGs and mutilated graphs differ.

The edges among random variables on the SWIG encode the factorization of the joint distribution $p(V(a))$.
More precisely, the FFRCISTG model (and thus the NPSEM-IE) imply that for any $A \subseteq V$, and $a \in {\mathfrak X}_A$, the distribution $p(V(a))$ factorizes with respect to ${\cal G}(V(a))$.  In other words,
\begin{align}\label{eq:swig-dag-factor}
p(V(a)) = \prod_{i=1}^K p\!\left(V_i(a) \left|\; V_{\pa_i\!\setminus A}(a) \right.\right).
\end{align}

Fixed nodes do not occur in the conditioning sets for the terms in (\ref{eq:swig-dag-factor})
and thus the presence or absence of edges ($a_i \rightarrow V_i(a_j)$) from fixed nodes to random nodes in ${\cal G}(V(a))$ are not reflected in this expression 
(\ref{eq:swig-dag-factor}).
However, the fact that a random node is not a descendant of a fixed node does encode information about causal irrelevance.
Specifically, if there is no directed path from the fixed node $a_j$ to $V_i(a)$ then $V_i(a)=V_i(a_{A\setminus \{j\}})$, hence under minimal labeling
$a_j$ will not appear in the label for the vertex $V_i(a_{\an^*_i})$ in ${\cal G}(V(a))$.\footnote{
\label{foot:weaker-swigs} In the Appendix we briefly consider using the SWIG to make inferences about weaker causal models, including the agnostic causal model, and models in which the absence of a directed edge corresponds to the absence of a population-level direct effect. In the latter models, the equality $V_i(a) = V_i(a \cap \an^*_i)$ would no longer hold, and minimal labelings are constructed using a (possibly strict) edge super-graph of the graph used for the factorization (\ref{eq:swig-dag-factor}); see \citep[Section 7]{thomas13swig}.}
Thus, as noted earlier, by causal irrelevance, $V_i(a) = V_i(a_{\an^*_i})$, where $\an^*_i$ consists of the fixed vertices that are ancestors of $V_i(a)$ in ${\cal G}(V(a))$.
Thus (\ref{eq:swig-dag-factor}) may be expressed as:
\[
p(V(a)) = \prod_{i=1}^K p\!\left(V_i(a_{\an_i^*}) \left|\; \left\{V_j(a_{\an_j^*}), \hbox{ for } j\in  {\pa_i\!\setminus A}\right\} \right.\right).
\]

More generally, paths commencing with a fixed node but on which every other node is random also encode information about functional dependence.
A path $\pi$ in a SWIG ${\cal G}(a)$ is said to be {\it Markov relevant} if at most one endpoint is a fixed vertex, and every non-endpoint is random.
A Markov relevant path $\pi$ in ${\cal G}(a)$  is {\it d-connecting given $V_{Z}(a)$} if every collider on $\pi$ is an ancestor of a vertex in $V_Z(a)$ and every non-collider
on $\pi$ is not in $V_Z(a)$.

It follows directly from (\ref{eq:swig-dag-factor}) that if $V_X(a)$ is d-separated from $V_Y(a)$ given $V_Z(a)$ in ${\cal G}(a)$ then 
$V_X(a) \ci V_Y(a)\,|\, V_Z(a)$ in $p(V(a))$, so that d-separation relations among random variables encode conditional independence.
 In addition, the absence of any d-connecting path in ${\cal G}(V(a))$  between a fixed node $a_j$ and a set of random vertices
$V_Y(a)$, given a (possibly empty) set of random variables $V_Z(a)$, encodes that $p(V_Y(a)\,|\,V_Z(a))$ does not depend on the value of $a_j$.
Thus we allow d-separation statements of the form $(a_j \ci_d V_Y(a) \mid V_Z(a))_{{\cal G}(V(a))}$.\footnote{This represents an extension of the notion of d-separation in \cite{thomas13swig}. %
Our extension here consists only in allowing fixed vertices to appear in at most one side of a d-separation statement (not the conditioning set). The semantics for these extended d-separation statements are given in (\ref{eq:dsep-semantics}) below.} 
More generally, given three disjoint subsets $Y, X, Z \subseteq V$, where $Z$ may be empty, and a set $A' \subseteq  A$, then 
\begin{equation}\label{eq:dsep-swig}
(V_Y(a) \ci_d V_X(a),a_{A'}   \mid V_Z(a))_{{\cal G}(V(a))}
\end{equation}
 if in the SWIG ${\cal G}(V(a))$ there is no path d-connecting a random vertex $V_i(a)$ with $i\in X$ or a fixed vertex $a_j$ with $j \in A'$ to a random vertex in $V_j(a)$ with $j \in Y$ given $V_Z(a)$. Note that, by definition, fixed vertices may only arise on one side of a d-separation statement (\ref{eq:dsep-swig}).
Conversely, a possibly d-connecting path may only contain at most one fixed node in which case it is an endpoint vertex (thus, as in \cite{thomas13swig}, fixed nodes never arise as non-endpoint vertices on d-connecting paths).



Results for DAG models with fixed nodes \cite{richardson17nested} imply the following:
\begin{prop}[SWIG global Markov property] \label{prop:extended-cdag-markov}
\begin{itemize}
\item[]
\end{itemize}
\noindent
Under the FFRCISTG for ${\cal G}$, for every set $A$,
 disjoint sets of random vertices $V_X(a)$, $V_Y(a)$, $V_Z(a)$ and a set of fixed nodes $a_{A'}$, where $A' \subseteq A$,
\begin{align}
\text{if } (V_Y(a) \ci_d V_X(a), a_{A'}   \,|\, V_Z(a))_{{\cal G}(V(a))} & \text{ then, for some }f(\cdot),\nonumber\\
 p(V_Y(a) = v_Y \,|\, V_Z(a) = v_Z, V_X(a) = v_X) &= p(V_Y(a) = v_Y \,|\, V_Z(a) = v_Z)\label{eq:dsep-semantics}\\
 &=  f(v_Y,v_Z,a_{A \setminus A'}). \nonumber
\end{align}
\end{prop}

\begin{exa}
\label{exa:front}
Consider the global Markov property associated with the SWIG ${\cal G}(a)$ in Figure \ref{fig:front-door}(b), corresponding to the FFRCISTG
model 
shown in Figure \ref{fig:front-door} (a). Since $a$ is d-separated from $Y(a)$ given $M(a)$ in ${\cal G}(a)$,
\begin{align}
p(Y(a) = y \mid M(a) = m) = f(y,m). \label{eq:front-door-constraint}
\end{align}
Hence this distribution is not a function of $a$, even though $M(a)$ and $Y(a)$ are minimally labeled, so $M(a)\neq M(a')$ and $Y(a)\neq Y(a')$ for $a\neq a'$.  In addition, it is well known that in the FFRCISTG model corresponding to Figure \ref{fig:front-door} (a),
\begin{align*}
p(Y(a) = y \,|\, M(a) = m) = \sum_{a'} p(y \,|\, m, a') p(a'),
\end{align*}
which is not equal to $p(y \mid m)$ under the given model.
Hence, the function $f(v_Y,v_Z,a_{A \setminus A'})$ is not necessarily equal to the conditional distribution $p(V_Y(a_{A \setminus A'}) \,|\, V_Z(a_{A \setminus A'}))$.
\end{exa}

\begin{rmk}
Since, by construction,  all edges in ${\cal G}(V(a))$ are directed out of $a_j$, in the case where $Z$ is the empty set, there is a d-connecting path between $a_j$ and $V_i(a)$ if and only if $a_j$ is an ancestor of $V_i(a)$ in ${\cal G}(V(a))$; as noted above, this is automatically reflected with the minimal labeling of the vertices.
\end{rmk}



In (\ref{eq:front-door-constraint}) we see an example where $p(Y(a) \,|\, M(a))$ does not depend on $a$, even though $Y(a)$ and $M(a)$ are minimally labeled. One might wonder whether it is possible to have the converse situation whereby a conditional distribution {\it does} depend on a fixed vertex that is not present in any minimal label. The next Proposition shows that this cannot arise:

\begin{prop}
In a minimally labeled SWIG ${\cal G}(a)$, if a fixed vertex  $a_i$ is d-connected to $V_j(a_{\an_{j}^*})$ given $\{V_{k_1}(a_{\an_{k_1}^*}),\ldots ,V_{k_p}(a_{\an_{k_p}^*})\}$
then either $i\in {\an_{j}^*}$ or $i\in {\an_{k_s}^*}$ for some $s$. 
\end{prop}

In words, if a fixed vertex is d-connected by a path to a random vertex given some conditioning set, then the fixed vertex either appears in the minimal label
for the other endpoint, or a vertex in the conditioning set. This follows since if there is a d-connecting path on which $a_i$ is an endpoint then, since $a_i$ only has children in ${\cal G}(a)$, the path is directed out of $a_i$.
The conclusion then follows since if the path contains no colliders then $V_j(a_{\an_{j}^*})$ is a descendant of $a_i$; if the path contains a collider then 
$a_i$ is an ancestor of that collider, which, by definition of d-connection is itself an ancestor of a vertex in $V_{k_s}(a_{\an_{k_s}^*})$.

\subsection{Factorization associated with the SWIG global Markov property}

As noted above, the factorization (\ref{eq:swig-dag-factor}) corresponds solely to the induced subgraph of ${\cal G}(a)$ on the random vertices.
We now derive the factorization corresponding to the SWIG global Markov property. Consider a single term in  (\ref{eq:swig-dag-factor}):
\begin{align}
\notag
\MoveEqLeft{p(V_i(a) = v_i \,|\, V_{\pa_i\!\setminus A}(a) = v_{\pa_i\!\setminus A} )}\\
\notag
&= p(V_i(a, v_{\pa_i\! \setminus A}) = v_i \,|\, V_{\pa_i\!\setminus A}(a, v_{\pa_i\! \setminus A}) = v_{\pa_i\!\setminus A} )\\
\notag
&= p(V_i(a, v_{\pa_i\! \setminus A}) = v_i)\\
&= p(V_i(a_{\pa_i \cap A}, v_{\pa_i\! \setminus A}) = v_i)\label{eq:only-parents-of-a}\\
&= p(V_i(a_{\pa_i \cap A}, v_{\pa_i\! \setminus A}) = v_i\,|\, 
V_{\pa_i \cap A}(a_{\pa_i \cap A}, v_{\pa_i\! \setminus A})= a_{\pa_i \cap A}, \notag\\[-4pt]
&\kern110pt\quad\quad\quad\quad V_{\pa_i\! \setminus A}(a_{\pa_i \cap A}, v_{\pa_i\! \setminus A})=v_{\pa_i\! \setminus A}
)\notag\\[2pt]
\label{eqn:g-single-term}
&= p(V_i = v_i \,|\, V_{\pa_i \cap A}= a_{\pa_i \cap A}, V_{\pa_i\! \setminus A}=v_{\pa_i\! \setminus A}).
\end{align}
Here the first equality follows from consistency; the second follows from  (\ref{eq:swig-dag-factor}) for ${\cal G}(a, v_{\pa_i\! \setminus A})$.
The third equality follows from causal irrelevance since if we intervene on all the parents of $V_i$ then no other variables have a causal effect on $V_i$.
The fourth line follows from (\ref{eq:swig-dag-factor}) for  ${\cal G}(a_{\pa_i \cap A}, v_{\pa_i\! \setminus A})$.
The fifth line again follows from consistency. Thus we have:
 
 \begin{prop}\label{prop:swig-pva-identified}
  Under the FFRCISTG models associated with a graph $\G$ we have the following identification formula:
\begin{align}
p(V(a) = v)
&= \prod_{i = 1}^{K} p(V_i(a) = v_i \,|\, V_{\pa_i\!\setminus A}(a) = v_{\pa_i\!\setminus A} )\\
\label{eqn:g}
&= \prod_{i=1}^K p(v_i \,\,|\, a_{\pa_i \cap A}, v_{\pa_i\! \setminus A}).
\end{align}
Thus $p(V(a))$ is identified if all of the conditional distributions in Equation {\rm (\ref{eqn:g})} are identified.\footnote{This may not hold in the absence of positivity; see \citep[Section \ref{sec:intervention-mediation-theory}]{rrs21volume_mediation_arxiv} for further discussion.}
\end{prop}


Now consider a DAG ${\cal G}^*(V\cup A^*)$ containing disjoint sets of vertices $V$ and $A^*$, with the same set of edges as in ${\cal G}(a)$ under the natural correspondence: $V_i \Leftrightarrow V_i(a)$ and $A^*_i \Leftrightarrow a_i$. Then  (\ref{eqn:g}) corresponds syntactically to the (subset of) terms in the DAG factorization for ${\cal G}^*$ associated with the variables in $V$. This then  establishes the SWIG global Markov property via results on conditional graphs \citep{richardson17nested}.%
\footnote{For the sole purpose of establishing the SWIG global Markov property it is sufficient to show that 
$p(V_i(a) = v_i \,|\, V_{\pa_i\!\setminus A}(a) = v_{\pa_i\!\setminus A} )$ is not a function of the fixed nodes
that are not in $\pa_i$, that is $a_{A\setminus \pa_i}$. This is established by (\ref{eq:only-parents-of-a}).  Under the FFRCISTG, 
$p(V_i(a_{\pa_i \cap A}, v_{\pa_i\! \setminus A}))$ exists even if $p( a_{\pa_i \cap A}, v_{\pa_i\! \setminus A})=0$.}

The modified factorization (\ref{eqn:g}) is known as the \emph{extended g-formula}
\citep{thomas13swig, robins04effects}.  Like the original factorization (\ref{eqn:d}), Equation (\ref{eqn:g}) has a term for every $V_i \in V$ not merely for every $V_i \in V \setminus A$.\footnote{
This is because the extended g-formula includes the value a variable takes on just before it is intervened upon and set to a constant $a_i$.}
  An alternative proof of the extended g-formula is given in \citep{thomas13swig}.


The following Proposition follows directly from Equation (\ref{eqn:g}) and is included here because a generalization of this result, Proposition \ref{prop:split} below, plays an important role in the identification of causal effects in DAGs with hidden variables.

\begin{prop}\label{prop:splitdag}
If ${\cal G}$ is a DAG with SWIG ${\cal G}(a)$ then for all $c_k \in {\mathfrak X}_k$
\begin{align*}
\MoveEqLeft{p(V\!(a,c_k)=v)}\\[-8pt]
&=p\!\left(V_{-k}(a)=v_{-k}, V_k(a)=c_k\right) \times 
\frac{\rule[-10pt]{0pt}{12pt} p\!\left(V_k(a)=v_k \,\left|\, V_{\pa_k^{{\cal G}(a)}}(a)=v_{\pa_k^{{\cal G}(a)}}\right. \right)}%
{\rule{0pt}{14pt} p\!\left(V_k(a) = c_k \,\left|\, V_{\pa_k^{{\cal G}(a)}}(a)= v_{\pa_k^{{\cal G}(a)}}\right.\right)},%
\end{align*}
where $V_{-k} \equiv \{V_1,\ldots, V_{k-1},V_{k+1},\ldots , V_K\}$, provided the conditional probability in the denominator is positive.
\end{prop}

\subsection{SWIG representation of the defining FFRCISTG assumptions}

Consider the special case in which $A=V$; in the resulting graph ${\cal G}(V(v))$ every variable (in $V$) has been split and thus no pair of random variables are joined by an edge. The factorization (\ref{eq:swig-dag-factor}) then becomes:
\begin{align}\label{eq:swig-all-split-factor}
p(V(v^*)=v) = \prod_{i=1}^K p\!\left(V_i(v^*)=v_i \right) =  \prod_{i=1}^K p\!\left(V_i(v^*_{\pa_i})=v_i \right),
\end{align}
and thus for a fixed $v^* \in {\mathfrak X}_V$ the one-step ahead counterfactuals $V_1(v^*_{\pa_1}),\ldots , V_K(v^*_{\pa_K})$ are independent. 
Note that (\ref{eq:swig-all-split-factor}) holding for all $v^* \in {\mathfrak X}_V$ is equivalent to (\ref{eqn:swm}) and thus  defines the FFRCISTG model.


\begin{figure*}
	\begin{center}
		\begin{tikzpicture}[>=stealth, node distance=1.4cm]
		\tikzstyle{format} = [draw, thick, circle, minimum size=4.0mm,
		inner sep=1pt]
		\tikzstyle{unode} = [draw, thick, circle, minimum size=1.0mm,
		inner sep=0pt,outer sep=0.9pt]
		\tikzstyle{square} = [draw, very thick, rectangle, minimum size=4mm]

		\begin{scope}
		\path[->,  node distance =1.0cm, line width=0.9pt]
		node[format, shape=ellipse] (a) {$A$}
		node[format, shape=ellipse, right of=a, xshift=0.0cm] (m) {$M$}			
		node[format, shape=ellipse, right of=m, xshift=0.0cm,yshift=0.0cm] (y) {$Y$}
		node[format, color=gray, shape=ellipse, yshift=-0.2cm, above of=m] (h) {$H$}
		
		(a) edge[blue] (m)
		(m) edge[blue] (y)
		(h) edge[red] (a)
		(h) edge[red] (y)
		
		node[below of=a, yshift=0.4cm, xshift=1.0cm] (l) {$(a)$}
		;
		\end{scope}


		\begin{scope}[xshift=3.5cm]
		\begin{scope}
			\tikzset{line width=0.9pt, inner sep=1.8pt, swig vsplit={gap=6pt, inner line width right=0.3pt}}	
				\node[xshift=0.0cm, yshift=0.0cm, name=a, shape=swig vsplit]{
        					\nodepart{left}{$A$}
        					\nodepart{right}{$a$} };
		\end{scope}
		\path[->,  line width=0.9pt]
		node[format, shape=ellipse, right of=a, xshift=0.3cm] (m) {$M(a)$}			
		node[format, shape=ellipse, right of=m, xshift=0.3cm,yshift=0.0cm] (y) {$Y(a)$}
		node[format, color=gray, shape=ellipse, yshift=-0.5cm, above of=m] (h) {$H$}
		
		(a) edge[blue] (m)
		(m) edge[blue] (y)
		(h) edge[red, in=100, out=180] (a)
		(h) edge[red] (y)
		
		node[below of=a, yshift=0.8cm, xshift=1.7cm] (l) {$(b)$}
		;
		\end{scope}	
		
		\begin{scope}[xshift=8.4cm, yshift=0.0cm]
		\begin{scope}
			\tikzset{line width=0.9pt, inner sep=1.8pt, swig vsplit={gap=6pt, inner line width right=0.3pt}}	
				\node[xshift=0.0cm, yshift=0.0cm, name=a, shape=swig vsplit]{
        					\nodepart{left}{$A$}
        					\nodepart{right}{$a$} };
		\end{scope}
		\begin{scope}
			\tikzset{line width=0.9pt, inner sep=1.8pt, swig vsplit={gap=6pt, inner line width right=0.3pt}}	
				\node[xshift=0.8cm, yshift=0.0cm, right of=a, name=m, shape=swig vsplit]{
        					\nodepart{left}{$M(a)$}
        					\nodepart{right}{$m$} };
		\end{scope}
		\path[->,  line width=0.9pt]
		node[format, shape=ellipse, right of=m, xshift=0.9cm,yshift=0.0cm] (y) {$Y(m)$}
		node[format, color=gray, shape=ellipse, yshift=-0.5cm, above of=m] (h) {$H$}
		
		(a) edge[blue] (m)
		(m) edge[blue] (y)
		(h) edge[red, in=100, out=180] (a)
		(h) edge[red] (y)
		
		node[below of=a, yshift=0.8cm, xshift=2.3cm] (c) {$(c)$}
		;
		\end{scope}

		\begin{scope}[xshift=0.0cm, yshift=-2.0cm, node distance = 1.0cm]
		\path[->,  line width=0.9pt]
		node[format, shape=ellipse] (a) {$A$}
		node[format, shape=ellipse, right of=a, xshift=0.0cm] (m) {$M$}			
		node[format, shape=ellipse, right of=m, xshift=0.0cm,yshift=0.0cm] (y) {$Y$}
		
		(a) edge[blue] (m)
		(m) edge[blue] (y)
		(a) edge[<->, red, bend left] (y)
		
		node[below of=a, yshift=0.4cm, xshift=1.0cm] (l) {$(d)$}
		;
		\end{scope}


		\begin{scope}[xshift=3.5cm, yshift=-2.0cm]
		\begin{scope}
			\tikzset{line width=0.9pt, inner sep=1.8pt, swig vsplit={gap=6pt, inner line width right=0.3pt}}	
				\node[xshift=0.0cm, yshift=0.0cm, name=a, shape=swig vsplit]{
        					\nodepart{left}{$A$}
        					\nodepart{right}{$a$} };
		\end{scope}
		\path[->,  line width=0.9pt]
		node[format, shape=ellipse, right of=a, xshift=0.3cm] (m) {$M(a)$}			
		node[format, shape=ellipse, right of=m, xshift=0.3cm,yshift=0.0cm] (y) {$Y(a)$}
		
		(a) edge[blue] (m)
		(m) edge[blue] (y)
		
		(a) edge[<->, red, out=100] (y)
		
		node[below of=a, yshift=0.8cm, xshift=1.7cm] (l) {$(e)$}
		;
		\end{scope}

		\begin{scope}[xshift=9.0cm, yshift=-2.0cm]
		\begin{scope}
			\tikzset{line width=0.9pt, inner sep=2.4pt, swig vsplit={gap=6pt, inner line width right=0.3pt}}	
				\node[xshift=0.0cm, yshift=0.0cm, name=a, shape=swig vsplit]{
        					\nodepart{left}{$A$}
        					\nodepart{right}{$a$} };
		\end{scope}
		\path[->,  line width=0.9pt]
		node[format, shape=ellipse, right of=a, xshift=0.3cm] (m) {$M(a)$}			
		
		(a) edge[blue] (m)
		
		
		node[below of=a, yshift=0.8cm, xshift=1.7cm] (c) {$(f)$}
		;
		\end{scope}
		
		\end{tikzpicture}
	\end{center}
	\caption{(a) A hidden variable causal model.
		(b) A SWIG corresponding to an intervention that sets $A$ to $a$ in the causal model represented by (a).
		(c) A SWIG corresponding to an intervention that sets $A$ to $a$ and $M$ to $m$.
		(d) A latent projection of the DAG in (a).
		(e) A latent projection of the SWIG in (b).
		(f) A latent projection of the SWIG in (b) onto an ancestral set of vertices $A, M(a)$ and $a$.
	}
	\label{fig:front-door}
\end{figure*}
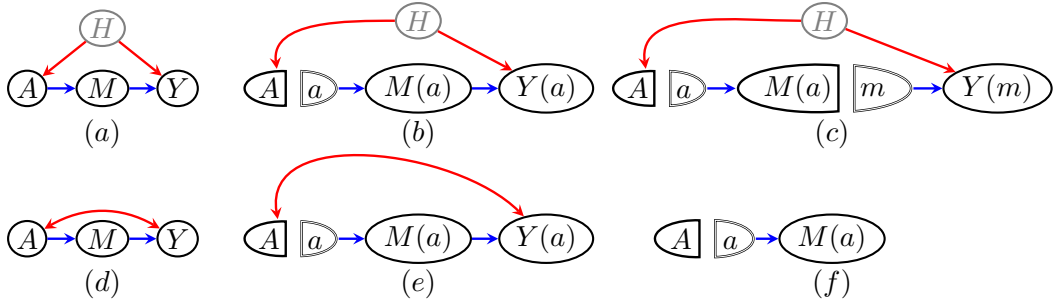

%

\section{The Potential Outcomes Calculus And Identification}
\label{sec:po-calc}

Pearl presented the three rules of {\it do}-calculus as an inference system for deriving identification results for causal inference problems.  The {\it do}-calculus is stated as three identities involving (conditional) interventional distributions, with preconditions given by d-separation (or m-separation) statements on graphs derived from the causal diagram ${\cal G}(V)$.  

Here we reformulate and extend these three rules as a  ``potential outcomes calculus'' or ``po-calculus'' for short.
The rules are as follows:
	\begin{align*}
	\text{\bf 1: }& p(Y(x) | Z(x),\! W(x)) \!=\! p(Y(x) | W(x))
	  \hspace{11.8mm} \text{ if } (Y(x) \ci Z(x) | W(x))_{{\cal G}(x)}, \\[6pt]
	\text{\bf 2: }& p(Y(x,\! z) | W(x,\! z)) \!=\! p(Y(x) | W(x),\! Z(x) \!=\! z) 
	  \hspace{1mm} \text{ if } (Y(x,\! z) \ci Z(x,\! z) | W(x,\! z))_{{\cal G}(x, z)},\\[6pt]
	 \text{\bf 3: }& p(Y(x,\! z)) \!=\! p(Y(x)) 
	 \hspace{36.7mm}\text{ if } (Y(x,z) \ci z)_{{\cal G}(x,z)},
	\end{align*}
where ${\cal G}(x)$ and ${\cal G}(x,z)$ are SWIGs describing interventions on $X$ and $X\cup Z$.
The sets $Z$, $Y$ and $W$ are assumed to be disjoint; $X$ may overlap with the other sets, but if $Z\cap X \neq \emptyset$ then we require $x_{X\cap Z} = z_{X\cap Z}$, so that the assignments are consistent.

Rule 1 can be viewed as the part of the SWIG global Markov property pertaining to random (rather than fixed) variables.

Rule 2 can be viewed as a kind of generalized conditional ignorability rule
which follows from Rule 1 and recursive substitution. Specifically   
by recursive substitution or minimal labelling  $Z(x,z)=Z(x)$; further,  
\begin{align*}
p(Y(x,z)\,|\,W(x,z))&=p(Y(x,z)\,|\,W(x,z),Z(x)=z)\\
&=p(Y(x)\,|\,W(x),Z(x)=z)
\end{align*}
here the first equality follows by the given d-separation and Rule 1, while the second follows from
 consistency (or recursive substitution) since $Y(x,z)=Y(x)$ and $W(x,z)=W(x)$ given that $Z(x)=z$.

Rule 3 expresses the property of causal irrelevance that interventions only affect descendants:
 Note that the Rule 3 condition $(Y(x,z) \ci z)_{{\cal G}(x,z)}$ is, by definition, equivalent to the fixed vertex (or vertices) $z$ not being an ancestor of any vertex in $Y(x,z)$
in the SWIG ${\cal G}(x,z)$ where the vertices in $X$ and $Z$ have been split.

Further, if a variable $Y(x)$ appears in the SWIG ${\cal G}(x,z)$ (with minimal labeling), then there is no directed path from any fixed vertex in $z$ to $Y(x)$, and $Y(x)=Y(x,z)$.\footnote{\label{foot:r3} Thus under the counterfactual model, as defined by one-step ahead counterfactuals, $V_i(a_{\pa_i})$ and recursive substitution (\ref{eqn:rec-sub}), we have a stronger implication than Rule 3: if $(Y(x,z) \ci z)_{{\cal G}(x,z)}$ then $Y(x,z) = Y(x)$.
Notwithstanding this, we formulate Rule 3 in terms of the equality of distributions because we wish these rules to be logically equivalent to the original {\it do}-calculus and also apply to weaker causal models; see Footnote \ref{foot:weaker-swigs} and Section \ref{sec:weaker}.}
Thus the minimal labeling of the SWIG implicitly encodes all applications of Rule 3, in the sense that if the (minimally labeled) vertex $Y(x)$ is present in the SWIG then for any set $Z$, disjoint from $X$, $p(Y(x)) = p(Y(x,z))$.


As we have shown here, the po-calculus directly follows from the SWIG global Markov property, which is implied by both the FFRCISTG model (and thus the NPSEM-IE), consistency, and causal irrelevance, where the latter two hold for any NPSEM.  

Rule 3, as stated here,\footnote{Rule 3 in this paper is called Rule $3^*$ by \cite{malinsky19po}.}  simply states that
interventions only affect descendants and thus is simpler than  Rule 3 in the original formulation of the {\it do}-calculus.
It is proved in \citep{malinsky19po} that this reformulated Rule 3, in conjunction with the other two rules, 
is equivalent to Pearl's {\it do}-calculus, in the sense that the three rules stated here imply the original three rules.
The rules stated here are more general in that we allow $X$ to overlap with $Y$, $Z$ and $W$, which is not possible within the framework and notation of the original {\it do}-calculus. As a consequence, as we will show below, there are additional identification results that follow from the po-calculus, but not the {\it do}-calculus. However, if we restrict the po-calculus rules to the case where $X$ does not overlap with $Y$ and $Z$ then they are equivalent to the {\it do}-calculus.

It may also be noted that the po-calculus is formulated using a uniform type of graph, the SWIG, for displaying the preconditions for each rule.\footnote{Whereas the original {\it do}-calculus involves three different constructions: $G_{\overline{X}}$, $G_{\overline{X}\underline{Z}}$ and
$G_{\overline{X(Z(W))}}$.}

\begin{rmk}
We note that there are other types of equality between distributions that do not correspond to a single application of the {\em po}-calculus rules.
For instance, in Example \ref{exa:front}, it follows that
\begin{equation}\label{eq:front-independence}
p(Y(a)\,|\,M(a))\; =\; p(Y(a')\,|\,M(a'))\quad\hbox{for } a, a' \in \mathfrak{X}_A.
\end{equation}
This holds even though $Y(a)$ and $M(a)$ depend on $a$ and $p(Y(a)\,|\,M(a))\,\neq\, p(Y\,|\,M)$. This is a form of independence.\footnote{Formally we may think of  $P(Y(a)\,|\,M(a))$ as forming a kernel $q(y\,|\,m,a)$, which is a set of conditional distributions indexed by $a$. The constraint is then an independence in this kernel; \citep{richardson17nested}.}
Such constraints are captured by the full global Markov property for SWIGs: notice that $a$ is d-separated from $Y(a)$ given $M(a)$ in the SWIG shown in Figure \ref{fig:front-door}(b). However, the equality in (\ref{eq:front-independence}) may be derived from three applications of the po-calculus (or the {\it do}-calculus) rules.
\end{rmk}





\section{Identification In Hidden Variable Causal Models}

If some variables in an NPSEM are unobserved, identification becomes more complicated, and some interventional distributions become non-identified.  Identification theory in NPSEMs associated with ${\cal G}(V \cup H)$, where $H$ are hidden variables is often described in terms of a special acyclic directed mixed graph (ADMG) ${\cal G}(V)$ obtained from ${\cal G}(V \cup H)$ via the \emph{latent projection} operation \citep{verma90equiv}.  Any two distinct hidden variable DAGs ${\cal G}_1(V \cup H_1)$, ${\cal G}_2(V \cup H_2)$ that share the latent projection ${\cal G}(V) = {\cal G}_1(V) = {\cal G}_2(V)$ also share all equality constraints on the observed marginal distribution \citep{evans:complete}, as well as non-parametric identification theory, in the sense that effects are identified in ${\cal G}_1$ if and only if they are identified in ${\cal G}_2$, and by the same functional \citep{richardson17nested}.

In cases where $p(V(a))$ is identified, the functional is a kind of modified factorization associated with nested Markov models of ADMGs \citep{richardson17nested}.

\subsection{Latent Projection ADMGs}
\label{subsec:latent}

Given a DAG ${\cal G}(V\cup{H})$, where ${V}$ are observed and ${H}$ are hidden variables, a {latent projection} ${\cal G}({V})$ is the following ADMG with a vertex set ${V}$. An edge $A \to B$ exists in ${\cal G}({V})$ if there exists a directed path from $A$ to $B$ in ${\cal G}({V}\cup{H})$ with all intermediate vertices in ${H}$.  Similarly, an edge $A \leftrightarrow B$ exists in ${\cal G}({V})$ if there exists a path without consecutive edges $\to \circ \gets$ from $A$ to $B$ with the first edge on the path of the form $A \gets$, the last edge on the path of the form $\to B$, and all intermediate vertices on the path in ${H}$.
Latent projections of hidden variable DAGs may be viewed as graphical versions of marginal distributions, in the following sense.
Just as conditional independences may be read off a DAG using d-separation, they may be read from an ADMG via the natural extension of d-separation to ADMGs which is called m-separation \cite{richardson03markov}.

If $p(V \cup H)$ factorizes with respect to ${\cal G}(V \cup H)$, then for any disjoint subsets $A,B,C$ of $V$, if $A$ is m-separated from $B$ given $C$, then $A$ is independent of $B$ conditionally on $C$ in the marginal distribution $p(V)$.  Since latent projections define an infinite class of hidden variable DAGs that share identification theory,
identification algorithms are typically defined directly on latent projections for simplicity.


Given $A \subseteq V$ in a hidden variable DAG ${\cal G}(V \cup H)$, we can construct the latent projection of the SWIG ${\cal G}(V(a) \cup H(a))$ directly from the ADMG ${\cal G}(V)$,  we denote the resulting ADMG (with fixed nodes) by ${\cal G}(V(a))$.
We can extend d-separation on SWIGs constructed from DAGs to m-separation on SWIGs constructed from ADMGs, and define the SWIG global Markov property on SWIG ADMGs analogously to the SWIG global Markov property on SWIG DAGs.  Similarly, we can restate po-calculus rules using m-separation on SWIG ADMGs.

As an example, the latent projection of the hidden variable DAG in Figure \ref{fig:front-door} (a) is shown in Figure \ref{fig:front-door} (d), while the latent projection of the SWIG in Figure \ref{fig:front-door} (b) is shown in Figure \ref{fig:front-door} (e).

All vertex relations defined in (\ref{eqn:vertex-rels}) translate without change to any SWIG ${\cal G}(V(a))$, except by convention $\dis_i^{{\cal G}(V(a))}$, $\mb_i^{{\cal G}(V(a))}$, and $\pre_i^{{\cal G}(V(a))}$ may only contain random vertices, in other words, they are subsets of $V(a)$.


We will describe a complete identification algorithm in hidden variable DAG models for all distributions of the form $p(Y(a))$, where $Y$ may potentially intersect $A$.  The original formulation of the problem in \citep{tian02on,shpitser06id,richardson17nested} assumed $Y \cap A = \emptyset$, and yielded the \emph{ID algorithm}.

We call our version of the algorithm the \emph{extended ID algorithm}, by analogy with the \emph{extended g-formula} (\ref{eqn:g}).
The extended ID algorithm will be formulated using SWIGs defined on latent projection ADMGs of the underlying hidden variable DAG.
The algorithm will take advantage of the fact that under certain assumptions given by the causal model, a single splitting operation that defines a counterfactual distribution in a SWIG can be phrased in terms of the observed data distribution.  This insight can be applied inductively to obtain results of multiple splitting operations as functionals of the observed data distribution.

The extended ID algorithm expresses the functional for $p(Y(a))$ as a counterfactual factorization in a certain SWIG ADMG, where terms of the factorization correspond to districts in the SWIG.  It then aims to identify each term by finding a sequence of splitting operations, possibly interleaved with marginalization operations.  Perhaps surprisingly, this always suffices to obtain identification
whenever identification is possible.

\subsection{The Identified Splitting Operation In A SWIG}
\label{subsec:fix}

A general identification algorithm for interventional distributions in hidden variable DAG models involves, as an essential step,
expressing the counterfactual distribution $p(V(a,c_k))$ as a function of another counterfactual distributions $p(V(a))$, where one fewer variable ($V_k$) has been intervened on, using restrictions in ${\cal G}(V(a))$.  Specifically, we have the following generalization of Proposition \ref{prop:splitdag}:


\begin{prop}
\label{prop:split}
Given an ADMG ${\cal G}(V)$ with SWIG ${\cal G}(V(a))$, if $V_k(a)$ is not split, so $k\notin A$, and $V_k(a)$ is such that there is no other random vertex that is both a descendant of $V_k(a)$ and in the same district as $V_k(a)$ then for all $c_k \in {\mathfrak X}_k$:
\begin{align*}
\MoveEqLeft{p(V\!(a,c_k)=v)}\\[-8pt]
&=p\!\left(V_{-k}(a)=v_{-k}, V_k(a)=c_k\right) \times 
\frac{\rule[-10pt]{0pt}{12pt} p\!\left(V_k(a)=v_k \,\left|\, V_{\mb_k^{\GVa}}(a)=v_{\mb_k^{\GVa}}\right. \right)}%
{\rule{0pt}{14pt} p\!\left(V_k(a) = c_k \,\left|\, V_{\mb_k^{\GVa}}(a)= v_{\mb_k^{\GVa}}\right.\right)},%
\end{align*}
where $V_{-k} \equiv \{V_1,\ldots, V_{k-1},V_{k+1},\ldots , V_K\}$, provided the conditional probability in the denominator is positive.
\end{prop}
In words, this Proposition states that if $V_k(a)$ satisfies the graphical condition in ${\cal G}(V(a))$ then $p(V\!(a,c_k))$, the joint distribution
 over all variables (including $A$ and $V_k$) resulting from intervening to set $A$ to $a$  and $V_k$ to $c_k$, may be obtained from $p(V(a))$ by evaluating at $V_k(a)=c_k$ and multiplying by a ratio of conditional densities for $V_k(a)$. 

The graphical condition may be interpreted as requiring that in the world where we have already intervened on $A$,  there 
is no sequence of variables between $V_k$ and any of its causal descendants such that there is an unmeasured confounder between each pair.

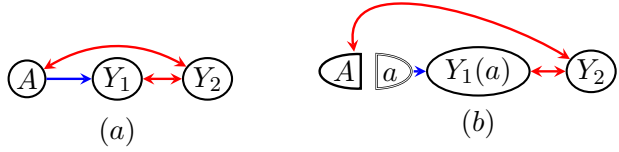
\begin{figure*}
	\begin{center}
		\begin{tikzpicture}[>=stealth, node distance=1.2cm]
		\tikzstyle{format} = [draw, thick, circle, minimum size=4.0mm,
		inner sep=1pt]
		\tikzstyle{unode} = [draw, thick, circle, minimum size=1.0mm,
		inner sep=0pt,outer sep=0.9pt]
		\tikzstyle{square} = [draw, very thick, rectangle, minimum size=4mm]

		\begin{scope}
		\path[->,  line width=0.9pt]
		node[format] (a) {$A$}
		node[format, shape=ellipse, right of=a, xshift=0.0cm] (y1) {$Y_1$}			
		node[format, shape=ellipse, right of=y1, xshift=0cm,yshift=0cm] (y2) {$Y_2$}

		(a) edge[blue] (y1)
		(a) edge[<->, red, bend left] (y2)
		(y1) edge[<->, red] (y2)

		node[below of=y1, yshift=0.5cm, xshift=0.0cm] (l) {$(a)$}
		;
		\end{scope}

		\begin{scope}[xshift=4.0cm]
		\begin{scope}
			\tikzset{line width=0.9pt, inner sep=1.8pt, swig vsplit={gap=6pt, inner line width right=0.3pt}}	
				\node[xshift=0.5cm, yshift=0.1cm, name=a, shape=swig vsplit]{
        					\nodepart{left}{$A$}
        					\nodepart{right}{$a$} };
		\end{scope}
		\path[->,  line width=0.9pt]
		node[format, shape=ellipse, right of=a, xshift=0.3cm] (y1) {$Y_1(a)$}			
		node[format, shape=ellipse, right of=y1, xshift=0.3cm,yshift=0cm] (y2) {$Y_2$}

		(a) edge[blue] (y1)
		(a) edge[<->, red, bend left, out=100] (y2)
		(y1) edge[<->, red] (y2)

		node[below of=y1, yshift=0.5cm, xshift=0.0cm] (l) {$(b)$}
		;
		\end{scope}
		
		\end{tikzpicture}
	\end{center}
	\caption{(a) A graph where $p(Y_1(a))$ and $p(Y_2(a))$ are identified, but Proposition \ref{prop:split} may not be applied.
	(b) A SWIG showing a splitting operation that is not identified according to Proposition \ref{prop:split}.
	}
	\label{fig:m-bias}
\end{figure*}

There exist counterfactual distributions which are identified, but where the above Proposition does not directly apply to the observed data distribution.  For example, in the graph in Figure \ref{fig:m-bias} (a), $p(Y_1(a)) = p(Y_1 \mid a)$, and $p(Y_2(a)) = p(Y_2)$.  Nevertheless, the preconditions to applying Proposition \ref{prop:split} do not apply to the original graph, meaning that the distribution represented by the SWIG in Figure \ref{fig:m-bias} (b) is not equal to the functional of the observed data distribution described in the Proposition.  In fact, the joint distribution associated with this SWIG is not identified at all, as was shown in \cite{tian02on}.  Nevertheless, identification of $p(Y_2(a))$ and $p(Y_1(a))$ may be obtained by the identification algorithm we describe below.

In the next section we will apply the Proposition iteratively, in conjunction with marginalization steps, in order to obtain a complete algorithm for identifying a margin $p(Y(a))$.\\

\begin{prf}
Fix an ordering $\prec^\prime$ on vertex indices 
such that $\prec^\prime$ is topological in ${\cal G}(V(a))$ and
 such that no element in the district of $V_k(a)$ 
occurs later in the ordering than $V_k(a)$. Such a topological ordering exists because, by hypothesis, no vertex in the district of $V_k(a)$ is a descendant of $V_k(a)$.
For any index $j$, define $\pre^\prime_j$ to be the set of predecessor indices according to $\prec^\prime$.



\begin{align*}
\MoveEqLeft{p(V(a,c_k)=v)}\\
=&\prod_{i \in \pre^\prime_k} p(V_i(a,c_k)=v_i \,|\, V_{\pre^\prime_i}(a,c_k)=v_{\pre^\prime_i} )
\times p(V_k(a,c_k) =v_k\,|\, V_{\pre^\prime_k}(a,c_k) = v_{\pre^\prime_k})
\\[-6pt]
&
\kern30pt\times\!\!\!\!\!  \prod_{j \notin \{ k \} \cup \pre^\prime_k}\!\!\!\!\! p(V_j(a,c_k)=v_j \,|\, V_{\pre^\prime_j}(a,c_k)=v_{\pre^\prime_j} )
\\[8pt]
=& \prod_{i \in \pre^\prime_k} p(V_i(a)=v_i \,|\, V_{\pre^\prime_i}(a)=v_{\pre^\prime_i} ) \times p(V_k(a) =v_k\,|\, V_{\pre^\prime_k}(a) = v_{\pre^\prime_k})
\\[-6pt]
&
\kern30pt\times\!\!\!\!\!  \prod_{j \notin \{ k \} \cup \pre^\prime_k}\!\!\!\!\! p(V_j(a,c_k)=v_j \,|\, V_{\pre^\prime_j}(a,c_k)=v_{\pre^\prime_j} )
\end{align*}

\begin{align*}
=& \prod_{i \in \pre^\prime_k} p(V_i(a)=v_i \,|\, V_{\pre^\prime_i}(a)=v_{\pre^\prime_i}  ) \times p(V_k(a) =v_k\,|\, V_{\pre^\prime_k}(a) = v_{\pre^\prime_k})
\\[-6pt]
&
\kern30pt\times\!\!\!\!\!  \prod_{j \notin \{ k \} \cup \pre^\prime_k}\!\!\!\!\! p(V_j(a) = v_j \,|\, V_{\pre^\prime_j \setminus \{ k \}}(a)=v_{\pre^\prime_j}, V_k(a) = c_k )
\\[8pt]
=&\prod_{i \in \pre^\prime_k} p(V_i(a)=v_i \,|\, V_{\pre^\prime_i}(a)=v_{\pre^\prime_i} ) \times p(V_k(a) =c_k\,|\, V_{\pre^\prime_k}(a) = v_{\pre^\prime_k})
\\[-6pt]
&
\kern30pt\times\!\!\!\!\!  \prod_{j \notin \{ k \} \cup \pre^\prime_k}\!\!\!\!\! p(V_j(a) = v_j \,|\, V_{\pre^\prime_j \setminus \{ k \}}(a)=v_{\pre^\prime_j}, V_k(a) = c_k )
\\
&\kern30pt\times \left.{p(V_k(a)=v_k \,|\, V_{\pre^\prime_k}(a)=v_{\pre^\prime_k})}\right/\!
{p(V_k(a) = c_k \,|\, V_{\pre^\prime_k}(a)= v_{\pre^\prime_k})}\\[8pt]
=&\;\; p\!\left(V_{-k}(a)=v_{-k}, V_k(a)=c_k\right)\\
&\kern30pt\times \left.{p(V_k(a)=v_k \,|\, V_{\pre^\prime_k}(a)=v_{\pre^\prime_k})}\right/\!
{p(V_k(a) = c_k \,|\, V_{\pre^\prime_k}(a)= v_{\pre^\prime_k})}\\
=&\;\; p\!\left(V_{-k}(a)=v_{-k}, V_k(a)=c_k\right)
%
\frac{\rule[-9pt]{0pt}{12pt} p\!\left(V_k(a)=v_k \,\left|\, V_{\mb_k^{\GVa}}(a)=v_{\mb_k^{\GVa}}\right. \right)}%
{\rule{0pt}{13pt} p\!\left(V_k(a) = c_k \,\left|\, V_{\mb_k^{\GVa}}(a)= v_{\mb_k^{\GVa}}\right.\right)}.%
\end{align*}
Here the first identity is via the chain rule of probability applied to $p(V(a,c_k))$ using the ordering $\prec^\prime$,
the second by Rule 3 (causal irrelevance) applied to elements indexed by $\{ k \} \cup \pre^\prime_k$ in ${\cal G}(V(a,c_k))$,
the third by Rule 2 (generalized ignorability) applied to every term in the second product in ${\cal G}(V(a,c_k))$,
 and the assumption on $\prec^\prime$ that all elements of $\dis_k^{{\cal G}(V(a))}$ are in $V_{\pre^\prime_k}(a)$,
the fourth by multiplying and dividing by $p(V_k(a)\!=\!c_k\,|\, V_{\pre^\prime_k}(a)\!=\!v_{\pre^\prime_k})$, the fifth by the chain rule, the sixth by Rule 1 (m-separation) applied to ${\cal G}(V(a))$ and the definition of $\mb_k^{\GVa}$.
\end{prf}
\\


\subsection{The Extended ID Algorithm}
\label{subsec:id}

There are SWIGs ${\cal G}(V(a))$ for which, for some variable $V_k(a)$ we are not able to apply Proposition \ref{prop:split}, but where it may be applied to a SWIG ${\cal G}(Y(a))$, where $Y(a)$ is an ancestral subset of $V(a)$ in ${\cal G}(V(a))$.  Here a set $Y$ of vertices in a (SWIG) ADMG ${\cal G}^*$ is said to be \emph{ancestral} if $V_i \in Y$ implies $\an_i^{{\cal G}^*} \subseteq Y$.

Marginal distributions $p(Y(a))$ obtained from $p(V(a))$ that correspond to ancestral sets in ${\cal G}(V(a))$ have the nice property that a latent projection ${\cal G}(Y(a))$ is always equal to an induced subgraph $({\cal G}(V(a)))_{Y(a)}$ of a SWIG ${\cal G}(V(a))$, with ${\cal G}(Y(a))$ having strictly fewer vertices and edges than ${\cal G}(V(a))$ if $Y(a) \subset V(a)$.  For example, given the SWIG in Figure \ref{fig:triangle} (e), the latent projection onto the ancestral subset $A$, $M(a)$ and $a$ yields the SWIG shown in Figure \ref{fig:triangle} (f).
We describe the precise way in which splitting and ancestral margin operations are used to obtain identification below.

Specifically, complete non-parametric identification for intervention distributions associated with the FFRCISTG model may be obtained from: (i) the district factorization in the appropriate SWIG, (ii) the identified splitting operation described in the previous section, and (iii) marginalization steps that lead to marginal distributions corresponding to ancestral sets of vertices in SWIGs. All of these steps may be justified via the po-calculus.

For any (possibly intersecting) subsets ${Y},{A}$ of ${V}$ in a latent projection $\G({V})$ representing a causal DAG $\G({V}\cup{H})$,
define ${Y}^*(a)$ to be the random ancestors of $Y(a)$ in $\G({V}(a))$. 
Clearly, if $p(Y^*(a))$ is identified, then we may recover $p(Y(a))$ since:
\begin{equation}
P(Y(a)=y) = \sum_{u \in {\mathfrak X}_{Y^*\setminus Y}} p(V_{Y}(a) = y, V_{Y^*\setminus Y}(a)=u).
\end{equation}
Though less obvious, extensions of results in \citep{shpitser06id} imply that the converse also holds, so that if 
$p(Y(a))$ is identified (for all parameter values) then $p(Y^*(a))$ is identified.
Consequently, in the foregoing we will assume that $Y(a)$ is an ancestral set of (random) vertices in $\G({V}(a))$.

If $p(Y(a))$ is identified then this may be obtained by breaking this joint distribution into districts in $\G({Y}(a))$.  
For each such district $D(a)$, define the set of \emph{strict (random) parents} as
$\pas_{D(a)}^{\GYa} \equiv \pa^{\GYa}_{D(a)} \setminus (D(a) \cup a)$.


First, we show that $p(Y(a) = y)$ can be factorized into a set of terms of the form $p(D(a,v_{\pas_{D(a)}^{\GYa}}))$, as follows.
\begin{align}
\notag
\MoveEqLeft{p(Y(a) = v_Y)}\\ 
=& 
\prod_{i \in Y} p\!\left(\!V_i(a) = v_i \,\left|\,
V_{Y \cap \pre_i}(a)= v_{Y \cap \pre_i}
\!\right.\right)\\
\notag
=& 
\prod_{D \in {\cal D}(\GYa)} \prod_{i \in D} 
p\!\left(\!V_i(a) = v_i \,\left|\,
V_{Y \cap \pre_i}(a)= v_{Y \cap \pre_i}
%
\!\right.\right)\\
\notag
=&  \prod_{D \in {\cal D}(\GYa)} \prod_{i \in D} p(V_i(a,v_{\pas^{\GYa}_D})=v_i\,|\, V_{D \cap  \pre_i}(a,v_{\pas^{\GYa}_D})=v_{D \cap  \pre_i})\\
=& \prod_{D \in {\cal D}(\GYa)} p\!\left(\!V_{D}(a,v_{\pas_{D}^{\GYa}}) = v_{D}\!\right).
\label{eqn:line-4}
\end{align}
Here the first two lines follow by the chain rule of probability, term grouping, and the fact that in any ADMG, including a SWIG ADMG, the set of districts partitions the set of random vertices.  The third equality follows because of the following:
\begin{align}
\MoveEqLeft[6]{p(V_i(a)=v_i\,|\, V_{Y\cap \pre_i}(a)=v_{Y\cap \pre_i})}\notag\\
&= p\!\left(V_i(a)=v_i \,\middle|\, V_{\mb^{\GYa}_i\cap \pre_i}(a)=v_{\mb^{\GYa}_i\cap \pre_i}\right)\label{eqn:2nd-line}\\
&= p\!\left(\vphantom{V_{\pas^{\GYa}_D \cap \pre_i}} V_i(a)=v_i \,\middle|\, V_{D \cap  \pre_i}(a)=v_{D \cap  \pre_i},  
V_{\pas^{\GYa}_D \cap \pre_i}(a)=v_{\pas^{\GYa}_D \cap \pre_i}\right)\notag\\[2pt]
&= p(V_i(a,b_i) \,|\, V_{D\cap  \pre_i}(a,b_i)=v_{D \cap  \pre_i})\notag\\
&= p\!\left(V_i(a,v_{\pas^{\GYa}_D})\,\middle|\, V_{D \cap  \pre_i}(a,v_{\pas^{\GYa}_D})=v_{D \cap  \pre_i}\right)\\[-16pt]
\notag
\end{align}
where $b_i=v_{\pas^{\GYa}_D \cap \pre_i}$, and $D= \dis^{\GYa}_i$. Here the first equality follows from Rule 1;\footnote{Note that the Markov blanket of $i$ in the subgraph of ${\cal G}(Y(a))$ restricted to predecessors of $i$ is, in general, a strict subset of the predecessors of $i$ in the Markov blanket of $i$ in ${\cal G}(Y(a))$.  Consequently, the conditioning set is in the terms of (\ref{eqn:2nd-line}) may not be minimal.}
 the second follows from the definition of the Markov blanket  of $V_i(a)$ in ${\cal G}(Y(a))$;
the third follows from Rule 2 since $V_i(a,b_i)$ is m-separated from $B_i(a,b_i) \equiv V_{\pas^{\GYa}_D \cap \pre_i}(a,b_i)$ in $\G(a,b_i)$;
the fourth is an application of Rule 3 since vertices in $V_{\pas^{\GYa}_D \setminus \pre_i}$ are ordered after $V_i$ and hence are not ancestors of $V_i$ in $\G$, and thus also in $\G(a,b_i)$.

Next, we consider whether each term of the form $p(V_{D}(a,v_{\pas_{D}^{\GYa}}))$ is identified from $p(V)$ by inductively applying the identified splitting operation in Proposition \ref{prop:split} to every element $V_j$ in $A \cup (V \setminus D)$ in a sequence such that the precondition of Proposition \ref{prop:split} is satisfied at every step, and marginalizing $V_j(a)$ at every step whenever 
$V_j\notin D$. (Hence $V_j$ will be split unless $V_j \in D\setminus A$.)
$p(Y(a))$ is identified if for every district $D \in {\cal D}({\cal G}(Y(a)))$, the corresponding term $p(V_{D}(a,v_{\pas_{D}^{\GYa}}))$ is identified in this way.
In fact, the above method of identification is sufficient and necessary for identification of $p(Y(a))$.
See the Appendix for details.

For the special case where $Y\cap A= \emptyset$ the resulting identified functionals were first described as an algorithm in \cite{tian02on}, and proven to be complete in \cite{huang06do,shpitser06id}. In both versions of the algorithm, the identifiable terms corresponding to districts $D(a)$ in ${\cal G}(Y(a))$ form parts of the \emph{nested Markov factorization}
of an ADMG, and the algorithm may thus be viewed as giving a modified nested factorization of an ADMG, just as the extended g-formula is a modified DAG factorization.  For more details, see \cite{richardson17nested}.


\subsection{Identification Of Conditional Interventional Distributions}

Targets of inference in causal inference are often functions of \emph{conditional} counterfactual distributions $p(Y(a) | Z(a))$ rather than marginal distributions $p(Y(a))$.  Such targets arise, for instance, when effects within certain subgroups are of interest, or when investigating relationships between primary and secondary outcomes.  A straightforward modification of the above algorithm yields identification in such settings.

Fix $Y,Z,A$ where $Y,Z$ are disjoint, but may both intersect $A$.  Fix the largest subset $W \subseteq Z$, with $Z' = Z \setminus W$, such that $Z'(a,z')$ is m-separated from $Y(a,z')$ given $W(a,z')$ in ${\cal G}(V(a,z'))$.  Then, by rule 2, $p(Y(a) \,|\, W(a), Z'(a) = z') = p(Y(a,z') \,|\, W(a,z'))$.
Next, let $A'$ be a maximal subset of $Z \cap A$ such that $A'(a,z') \ci Y(a,z') \mid \{ W(a,z') : W \in Z \setminus (Z' \cup A') \}$.
Then $p(Y(a) \,|\, Z(a))$ is identified if $p(Y(a,z'), \{ W(a,z') : W \in Z \setminus (Z' \cup A') \})$ is identified.  In fact, we have:
\begin{align*}
p(Y(a) \,|\, Z(a)) =
\left.
\frac{
p(Y(a,z'), \{ W(a,z') : W \in Z \setminus (Z' \cup A') \})
}{
p(\{ W(a,z') : W \in Z \setminus (Z' \cup A') \})
} \right|_{Z \setminus A' = z_{Z \setminus A'}}.
\end{align*}
As we show in the Appendix, this condition is also necessary since if $p(Y(a,z'), \{ W(a,z') : W \in Z \setminus (Z' \cup A') \})$ is not identified,
$p(Y(a) \,|\, Z(a))$ is also not identified.

\subsection{Representing context specific independence using SWIGs}\label{subsec:ivermectin}

We now discuss an extension of SWIGs due to 
 \citet{dahabreh2019generalizing} and \citet{sarvet:graphical:2020} that demonstrates that SWIGs have greater expressive power than
standard causal DAGs because of their ability to represent  interventional context
specific conditional independence. 

\begin{figure}
	\begin{center}
		\begin{tikzpicture}[>=stealth, node distance=1.2cm]
		\tikzstyle{format} = [draw, thick, circle, minimum size=4.0mm,
		inner sep=1pt]
		\tikzstyle{unode} = [draw, thick, circle, minimum size=1.0mm,
		inner sep=0pt,outer sep=0.9pt]
		\tikzstyle{square} = [draw, very thick, rectangle, minimum size=4mm]

	\begin{scope}[xshift=0cm]
		\path[->,  line width=0.9pt]
		node[format, shape=ellipse] (a) {$A$}
		node[format, shape=ellipse, above right of=a, yshift=-0.3cm, fill=lightgray] (s) {$S$}			
		node[format, shape=ellipse, below right of=a, yshift=+0.3cm] (m) {$M$}
		
		node[format, shape=ellipse, left= of s, xshift=0.25cm, fill=lightgray] (u) {$U$}

		node[format, shape=ellipse, right of=s,xshift=-0.2cm, fill=lightgray] (r) {$R$}
		node[format, shape=ellipse, right of=m,xshift=-0.2cm] (y) {$Y$}

		(a) edge[blue] (s)
		(a) edge[blue] (m)
		(s) edge[blue] (r)
		(m) edge[blue] (r)
		(m) edge[blue] (y)
		(r) edge[blue] (y)
		
		(u) edge[blue, bend left] (r)
		(u) edge[blue, out=270,in=180] (m)

		node[below of=m]{(a)}	
		;
	\end{scope}

	\begin{scope}[xshift=6.1cm]
		\path[->,  line width=0.9pt]
		node[format, shape=ellipse] (a) {$A$}
		node[ shape=ellipse, above right of=a, yshift=-0.3cm] (s) {}			
		node[format, shape=ellipse, below right of=a, yshift=+0.3cm] (m) {$M$}
		
		node[ shape=ellipse, left= of s, xshift=0.25cm] (u) {}

		node[ shape=ellipse, right of=s,xshift=-0.2cm] (r) {}
		node[format, shape=ellipse, right of=m,xshift=-0.2cm] (y) {$Y$}

		(a) edge[blue] (m)
		(m) edge[blue] (y)
		

		(a) edge[blue, bend left] (y)
		(m) edge[<->, bend right, red] (y)

		node[below of=m]{(a$^*$)}	
		;
		\end{scope}
		
	\begin{scope}[yshift=-3.5cm,xshift=-0.6cm]

		\begin{scope}
			\tikzset{line width=0.9pt, inner sep=1.8pt, swig vsplit={gap=6pt, inner line width right=0.3pt}}	
				\node[ xshift=0.0cm, yshift=0.0cm, name=a, shape=swig vsplit]{
        					\nodepart{left}{$A$}
        					\nodepart{right}{$a$} };
		\end{scope}

		\path[->,  line width=0.9pt]
		node[format, shape=ellipse, above right of=a, xshift=0.5cm, yshift=-0.3cm, fill=lightgray] (s) {$S(a)$}			
		;

		\begin{scope}
			\tikzset{line width=0.9pt, inner sep=1.8pt, swig vsplit={gap=6pt, inner line width right=0.3pt}}	
				\node[below right of=a, xshift=0.7cm, yshift=+0.3cm, name=m, shape=swig vsplit]{
        					\nodepart{left}{$M(a)$}
        					\nodepart{right}{$m$} };
		\end{scope}

		\path[->,  line width=0.9pt]

		node[format, shape=ellipse, left= of s, xshift=-0.1cm, fill=lightgray] (u) {$U$}

		node[format, shape=ellipse, right of=s, xshift=0.8cm, fill=lightgray] (r) {$R(a,m)$}
		node[format, shape=ellipse, right of=m, xshift=1.2cm] (y) {$Y(a,m)$}

		(a) edge[blue] (s)
		(a) edge[blue] (m)
		(s) edge[blue] (r)
		(m) edge[blue] (y)
		(r) edge[blue] (y)
		(m) edge[blue, out=20,in=240] (r)

		(u) edge[blue, out=270,in=180] (m)
		
		(u) edge[blue,bend left] (r)

		node[below of=m]{(b)}	
		;
		\end{scope}

	\begin{scope}[yshift=-3.5cm, xshift=5.5cm]

		\begin{scope}
			\tikzset{line width=0.9pt, inner sep=1.8pt, swig vsplit={gap=6pt, inner line width right=0.3pt}}	
				\node[ xshift=0.0cm, yshift=0.0cm, name=a, shape=swig vsplit]{
        					\nodepart{left}{$A$}
        					\nodepart{right}{$a$} };
		\end{scope}

		\path[->,  line width=0.9pt]
		node[ shape=ellipse, above right of=a, xshift=0.5cm, yshift=-0.3cm] (s) {}			
		;

		\begin{scope}
			\tikzset{line width=0.9pt, inner sep=1.8pt, swig vsplit={gap=6pt, inner line width right=0.3pt}}	
				\node[below right of=a, xshift=0.7cm, yshift=+0.3cm, name=m, shape=swig vsplit]{
        					\nodepart{left}{$M(a)$}
        					\nodepart{right}{$m$} };
		\end{scope}

		\path[->,  line width=0.9pt]
		node[ shape=ellipse, right of=s, xshift=0.4cm] (r) {}
		node[format, shape=ellipse, right of=m, xshift=1.2cm] (y) {$Y(a,m)$}

		(a) edge[blue] (m)
		(m) edge[blue] (y)
		
		(a) edge[blue, bend left] (y)
		
		(m) edge[<->, red, out=280, in=260, looseness=0.8] (y)

		node[below of=m]{(b*)}	
		;
		\end{scope}
		
	\begin{scope}[yshift=-7cm,xshift=-0.6cm]

		\begin{scope}
			\tikzset{line width=0.9pt, inner sep=1.8pt, swig vsplit={gap=6pt, inner line width right=0.3pt}}	
				\node[ xshift=0.0cm, yshift=0.0cm, name=a, shape=swig vsplit]{
        					\nodepart{left}{$A$}
        					\nodepart{right}{$a\!=\!0\,$} };
		\end{scope}

		\begin{scope}
			\tikzset{line width=0.9pt, inner sep=1.8pt, swig vsplit={gap=6pt, inner line width right=0.3pt}}	
				\node[below right of=a, xshift=0.7cm, yshift=+0.3cm, name=m, shape=swig vsplit]{
        					\nodepart{left}{$M(0)$}
        					\nodepart{right}{$m$} };
		\end{scope}
		
		\path[->,  line width=0.9pt]
		node[format, shape=ellipse, above right of=a, xshift=0.5cm, yshift=-0.3cm, fill=lightgray] (s) {$S(0)$}			

		node[format, shape=ellipse, left= of s, xshift=-0.15cm, fill=lightgray] (u) {$U$}

		node[format, shape=ellipse, right of=s, xshift=0.8cm, fill=lightgray] (r) {$R(0,m)$}
		node[format, shape=ellipse, right of=m, xshift=1.2cm] (y) {$Y(0,m)$}

		(a) edge[blue] (s)
		(a) edge[blue] (m)
		(s) edge[blue] (r)
		(m) edge[blue] (y)
		(r) edge[blue] (y)
		
		(u) edge[blue, out=220,in=200] (m)
		(m) edge[blue, out=20,in=240] (r)

		node[below of=m]{(c)}	
		;
		\end{scope}
		
	\begin{scope}[yshift=-7cm, xshift=5.5cm]

		\begin{scope}
			\tikzset{line width=0.9pt, inner sep=1.8pt, swig vsplit={gap=6pt, inner line width right=0.3pt}}	
				\node[ xshift=0.0cm, yshift=0.0cm, name=a, shape=swig vsplit]{
        					\nodepart{left}{$A$}
        					\nodepart{right}{$a\!=\!0\,$} };
		\end{scope}

		\begin{scope}
			\tikzset{line width=0.9pt, inner sep=1.8pt, swig vsplit={gap=6pt, inner line width right=0.3pt}}	
				\node[below right of=a, xshift=0.7cm, yshift=+0.3cm, name=m, shape=swig vsplit]{
        					\nodepart{left}{$M(0)$}
        					\nodepart{right}{$m$} };
		\end{scope}
		
		\path[->,  line width=0.9pt]
		node[ shape=ellipse, above right of=a, xshift=0.5cm, yshift=-0.3cm] (s) {}			


		node[ shape=ellipse, right of=s, xshift=0.3cm] (r) {}
		node[format, shape=ellipse, right of=m, xshift=1.2cm] (y) {$Y(0,m)$}

		(a) edge[blue] (m)
		(m) edge[blue] (y)
		
		(a) edge[blue, bend left] (y)

		;
		\end{scope}
		
		\begin{scope}
		
		\begin{scope}
			\tikzset{line width=0.9pt, inner sep=1.8pt, swig vsplit={gap=6pt, inner line width right=0.3pt}}	
				\node[below right of=a, xshift=0.7cm, yshift=+0.3cm, name=m, shape=swig vsplit]{
        					\nodepart{left}{$M(0)$}
        					\nodepart{right}{$m$} };
		\end{scope}
		
		\path[->,  line width=0.9pt]


		node[format, shape=ellipse, right of=m, xshift=1.2cm] (y) {$Y(0,m)$}

		(a) edge[blue] (m)
		(m) edge[blue] (y)
		

		node[below of=m]{(c*)}	
		;
		\end{scope}

	\begin{scope}[yshift=-10.5cm,xshift=-0.6cm]

		\begin{scope}
			\tikzset{line width=0.9pt, inner sep=1.8pt, swig vsplit={gap=6pt, inner line width right=0.3pt}}	
				\node[ xshift=0.0cm, yshift=0.0cm, name=a, shape=swig vsplit]{
        					\nodepart{left}{$A$}
        					\nodepart{right}{$a\!=\!1\,$} };
		\end{scope}

		\path[->,  line width=0.9pt]
		node[format, shape=ellipse, above right of=a, xshift=0.5cm, yshift=-0.3cm, fill=lightgray] (s) {$S(1)$}			
		;
		
		\begin{scope}
			\tikzset{line width=0.9pt, inner sep=1.8pt, swig vsplit={gap=6pt, inner line width right=0.3pt}}	
				\node[below right of=a, xshift=0.7cm, yshift=+0.3cm, name=m, shape=swig vsplit]{
        					\nodepart{left}{$M(1)$}
        					\nodepart{right}{$m$} };
		\end{scope}
		
		\path[->,  line width=0.9pt]
		
		node[format, shape=ellipse, left= of s, xshift=-0.1cm, fill=lightgray] (u) {$U$}

		node[format, shape=ellipse, right of=s, xshift=0.8cm, fill=lightgray] (r) {$R(1,m)$}
		node[format, shape=ellipse, right of=m, xshift=1.2cm] (y) {$Y(1,m)$}

		(a) edge[blue] (s)
		(a) edge[blue] (m)
		(s) edge[blue] (r)
		(m) edge[blue] (y)
		(r) edge[blue] (y)
		
		(u) edge[blue,bend left] (r)
		(m) edge[blue, out=20,in=240] (r)

		node[below of=m]{(d)}	
		;
	\end{scope}

		\begin{scope}[yshift=-10.5cm, xshift=5.5cm]

		\begin{scope}
			\tikzset{line width=0.9pt, inner sep=1.8pt, swig vsplit={gap=6pt, inner line width right=0.3pt}}	
				\node[ xshift=0.0cm, yshift=0.0cm, name=a, shape=swig vsplit]{
        					\nodepart{left}{$A$}
        					\nodepart{right}{$a\!=\!1\,$} };
		\end{scope}
		\begin{scope}
			\tikzset{line width=0.9pt, inner sep=1.8pt, swig vsplit={gap=6pt, inner line width right=0.3pt}}	
				\node[below right of=a, xshift=0.7cm, yshift=+0.3cm, name=m, shape=swig vsplit]{
        					\nodepart{left}{$M(1)$}
        					\nodepart{right}{$m$} };
		\end{scope}
		
		\path[->,  line width=0.9pt]


		node[format, shape=ellipse, right of=m, xshift=1.2cm] (y) {$Y(1,m)$}

		(a) edge[blue] (m)
		(m) edge[blue] (y)
		
		(a) edge[blue, bend left] (y)

		node[below of=m]{(d*)}	
		;
		\end{scope}

		\end{tikzpicture}
		\end{center}
\caption{
(a) A DAG $\cal G$ representing two studies of river blindness, described in
\citep[Section \protect\ref{subsec:testable-untestable}]{rrs21volume_mediation_arxiv}.
(b) The SWIG ${\cal G}(a,m)$ resulting from $\cal G$; (c) and (d) show SWIGs ${\cal G}(a=0,m)$ and
${\cal G}(a=1,m)$ that incorporate additional context specific causal information. (a$^*$), (b$^*$), (c$^*$), (d$^*$) show the corresponding latent projections.
\label{fig:torpedo}}
\end{figure}
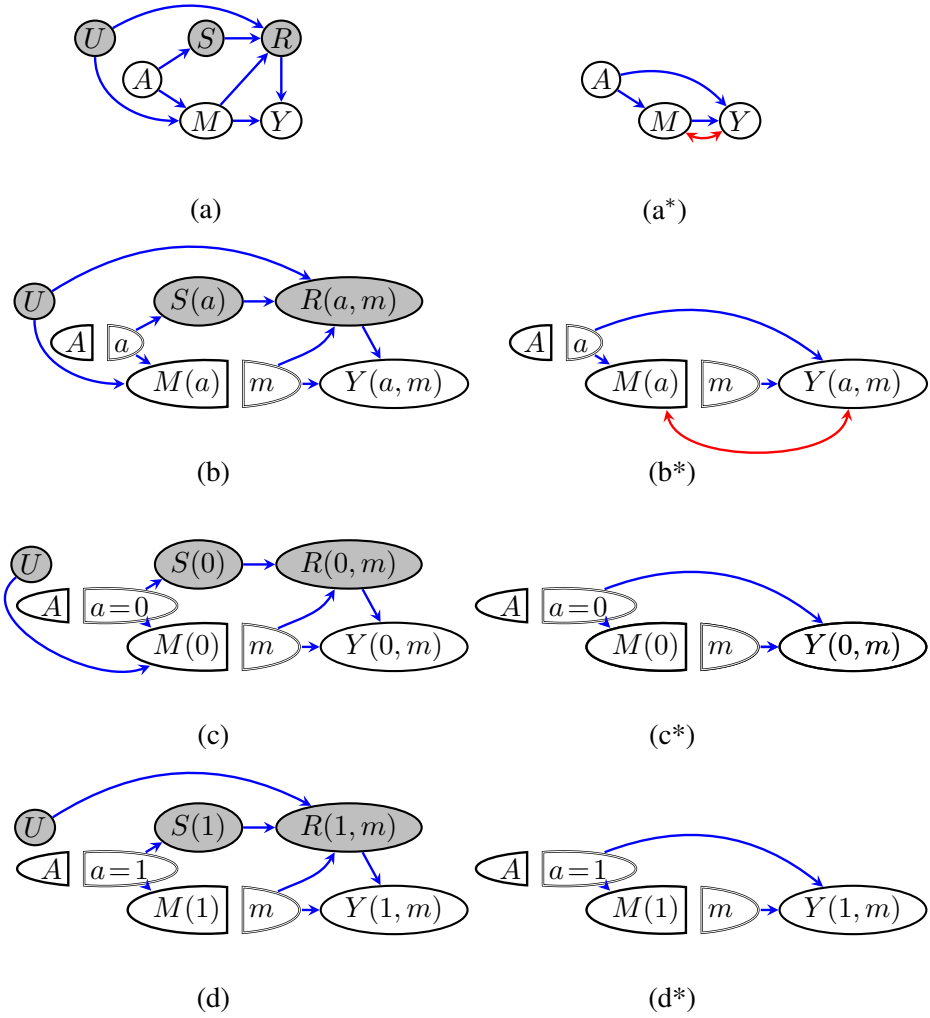

Consider the causal DAG shown in Figure \ref{fig:torpedo}(a)
where  $A$, $M$, $Y$ are observed and $U$, $R$, $S$ are unobserved. 
The latent projection
is given in Figure \ref{fig:torpedo}(a$^*$).
The SWIG resulting from a joint intervention setting $A$ to $a$ and $M$ to $m$
  is shown in Figure \ref{fig:torpedo}(b); the latent projection of this SWIG is shown in
Figure \ref{fig:torpedo}(b$^*$). The distribution of $Y(a,m)$ is not identified owing to the
presence of the edges $M \rightarrow Y \leftrightarrow M$ (also called a bow arc). 

However, suppose that, additional context specific 
subject matter knowledge\footnote{See the ivermectin study described in the companion paper \citep{rrs21volume_mediation_arxiv}.} implies that the following counterfactual
independences hold:
\[
U \ci R(a=0,m); \quad\quad  U\ci M(a=1).
\]
As a consequence, the edges $U\rightarrow R(0,m)$ in ${\cal G}(0,m)$ and
$U\rightarrow M(1)$ in ${\cal G}(1,m)$ may be removed, leading to the SWIGs shown in 
Figure \ref{fig:torpedo}(c) and (d), with corresponding latent projections shown
in Figure \ref{fig:torpedo}(c$^*$) and (d$^*$).


Applying d-separation to the latent projections in Figure \ref{fig:torpedo}(c$^*$) and (d$^*$) we see that\footnote{Recall that when testing d-separation in SWIGs, fixed nodes
such as $a=0$ in Figure \ref{fig:torpedo}(c$^*$) and $a=1$ in Figure \ref{fig:torpedo}(d$^*$) always block paths on which they occur as non-endpoint vertices.}
\begin{equation}\label{eq:yam-ind-ma-a}
Y(a,m) \ci M(a), A\quad \hbox{ for }a=0,1.
\end{equation}
Consequently, 
\begin{equation}\label{eq:am-intervention-identified}
P(Y\,|\,A=a,M=m) = P(Y(a,m)), 
\end{equation}
so that the joint effect of $A$ and $M$ on $Y$
 is identified for both $a=0$ and $a=1$. 

Given solely the DAG in Figure \ref{fig:torpedo} (a), with the latent projection in Figure \ref{fig:torpedo} (a$^*$), the equality (\ref{eq:am-intervention-identified}) would not be expected since it does not follow from existing methods
such as the {\it do}-calculus, the ID algorithm or the back-door criterion \cite{pearl09causality}, though see the recent work \cite{tikka:2019context-specific}. However, (\ref{eq:am-intervention-identified}) has a structural explanation in terms of the SWIGs corresponding to different treatment values of $A$.

In particular, the context-specific SWIG independences $U \ci R(0,m) \mid A, M(0)$ and $U {\ci}\, M(1) \mid A$, coupled with consistency,
 imply, respectively, the context specific independences $U \ci R \mid A=0$ and $U \ci M \mid A=1$ on the factual distribution.
These independences cannot be read off from the standard causal DAG shown in Figure \ref{fig:torpedo} (a).
This is because the absence of the $U\rightarrow M$ edge when $A$ is set to $1$ and the 
$U\rightarrow Y$ path when $A$ is set to $0$ are not represented in this DAG.

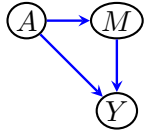
\begin{figure}
	\begin{center}
		\begin{tikzpicture}[>=stealth, node distance=1.2cm]
		\tikzstyle{format} = [draw, thick, circle, minimum size=4.0mm,
		inner sep=1pt]
		\tikzstyle{unode} = [draw, thick, circle, minimum size=1.0mm,
		inner sep=0pt,outer sep=0.9pt]
		\tikzstyle{square} = [draw, very thick, rectangle, minimum size=4mm]
		\begin{scope}
		\path[->,  line width=0.9pt]
		node[format, shape=ellipse] (a) {$A$}
		node[format, shape=ellipse, right of=a] (m) {$M$}			
		node[format, shape=ellipse, below of=m] (y) {$Y$}
		(a) edge[blue] (m)
		(m) edge[blue] (y)
		(a) edge[blue] (y)
		;
		\end{scope}
		\end{tikzpicture}
		\end{center}
\caption{A simple DAG containing a treatment $A$, an intermediate $M$ and a
response $Y$.
\label{fig:amyno-id}}
\end{figure}

Since, in addition to (\ref{eq:yam-ind-ma-a}), we also have $M(a) \ci A$ for $a=0,1$, it follows that the distribution of the counterfactuals $\{A, M(a), Y(a,m) \hbox{ for all } a,m \}$ obeys the FFRCISTG model associated with the graph shown in Figure \ref{fig:amyno-id} in which there are no bi-directed edges.
However, interestingly, the distribution of these counterfactuals does not obey the NPSEM-IE associated with Figure \ref{fig:amyno-id},
though the distribution does obey the NPSEM-IE (hence also the FFRCISTG) associated with Figure \ref{fig:torpedo} (a$^*$).%
\footnote{See \citep[Section \protect\ref{subsec:testable-untestable}]{rrs21volume_mediation_arxiv}.}

%

\section{Conclusion}
We wholeheartedly applaud Judea Pearl for his development and advocacy of graphical approaches to causal modeling. His approach represents a fundamental advance leading to many new insights and methods, {including complete identification theory for causal queries of all types, and extensions of d-separation to complex questions in causal modeling and missing data.}

{
\bibliography{references}
}

\section{Appendix}

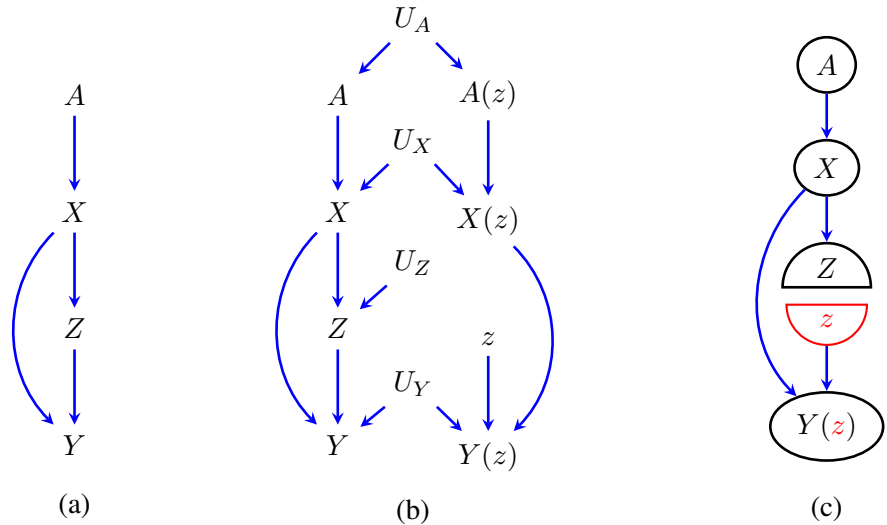
\begin{figure}
\begin{center}
\begin{tikzpicture}[
pre/.style={->,>=stealth,semithick,blue,line width = 1pt}]
\begin{scope}
\node[line width = 1pt,name=t] {
$A$
};
\node[line width = 1pt,below=of t, name=x] {
$X$
};
\node[line width = 1pt,below=of x, name=z] {
$Z$
};
\node[line width = 1pt,below=of z, name=y] {
$Y$
};
\node[below=0.5cm of y.center, name=i]{(a)};
\draw[pre, ->] (x) to (z);
\draw[pre, ->] (z) to (y);
\draw[pre, ->] (t) to (x);
\draw[pre, ->] (x) to[out=225,in=135] (y);
\end{scope}
\begin{scope}[xshift=3.5cm]
\begin{scope}
\node[line width = 1pt, name=t] {
$A$
};
\node[line width = 1pt,below=of t, name=x] {
$X$
};
\node[line width = 1pt,below=of x, name=z] {
$Z$
};
\node[line width = 1pt,below=of z, name=y] {
$Y$
};
\draw[pre, ->] (x) to (z);
\draw[pre, ->] (z) to (y);
\draw[pre, ->] (t) to (x);
\draw[pre, ->] (x) to[out=225,in=135] (y);
\end{scope}
\begin{scope}[xshift=1cm,yshift=1cm]
\node[line width = 1pt,name=ut] {
$U_A$
};
\node[line width = 1pt,below=of ut,name=ux] {
$U_X$
};
\node[line width = 1pt,below=of ux, name=uz] {
$U_Z$
};
\node[line width = 1pt,below=of uz, name=uy] {
$U_Y$
};
\end{scope}
\begin{scope}[xshift=2cm]
\node[line width = 1pt, name=ts] {
$A(z)$
};
\node[line width = 1pt, below=of ts, name=xs] {
$X(z)$
};
\node[line width = 1pt,below=of xs, name=zs] {
$z$
};
\node[line width = 1pt,below=of zs, name=ys] {
$Y(z)$
};
\draw[pre, ->] (zs) to (ys);
\draw[pre, ->] (ts) to (xs);
\draw[pre, ->] (xs) to[out=315,in=45] (ys);
\end{scope}
\draw[pre, ->] (ut) to (t);
\draw[pre, ->] (ut) to (ts);
\draw[pre, ->] (ux) to (x);
\draw[pre, ->] (ux) to (xs);
\draw[pre, ->] (uz) to (z);
\draw[pre, ->] (uy) to (y);
\draw[pre, ->] (uy) to (ys);
\node[below=1.4cm of uy.center, name=ii]{(b)};
\end{scope}
\begin{scope}[xshift=10cm,yshift=0.4cm]
\node[line width = 1pt, style=draw, shape=ellipse,name=t] {
$A$
};
\node[line width = 1pt,style=draw,shape=ellipse,below=6mm of t, name=x] {
$X$
};
\node[line width = 1pt,style=draw,shape=semicircle, below=6mm of x, name=z] {
$Z$
};
\node[thick,name=znew,color=red,shape=semicircle, shape border rotate=180, style={draw},text width=4mm,align=center,below=2mm of z] {
$z$
};
\node[line width = 1pt,style=draw,shape=ellipse,below=6mm of znew, name=y] {
$Y({\textcolor{red}{z}})$
};
\node[below=0.8cm of y.center, name=iii]{(c)};
\draw[pre, ->] (x) to (z);
\draw[pre, ->] (znew) to (y);
\draw[pre, ->] (t) to (x);
\draw[pre, ->] (x) to[out=225,in=135] (y);
\end{scope}
\end{tikzpicture}
\end{center}
\caption{(a) A DAG $\cal G$. (b) The twin-network arising from intervening to set $Z$ to $z$. 
(c) The SWIG ${\cal G}(z)$.
  \label{fig:second-torpedo}}
\end{figure}

\subsection{Incompleteness of d-separation in Twin Networks due to deterministic relations.}

{
Twin networks \citep{balke94countereval}
 are an alternative way to combine graphs and counterfactuals which allow some of the
counterfactual independence relations implied by the NPSEM-IE to be read off via d-separation; see also
\citep{shpitser07hierarchy}, \citep[Section 7.1.4]{pearl09causality}.
However, d-separation is not complete for Twin Networks \citep{thomas13swig} since, as a consequence of consistency, certain variables in a twin network may be deterministically related. Consequently, it is possible for there to be a d-connecting path in a Twin Network and yet the corresponding conditional independence holds for all distributions in the model.
}

{
To see a simple example, consider the DAG shown in 
Figure \ref{fig:second-torpedo}(a), with the Twin Network and SWIG associated with intervening to set $Z$ to $z$, shown in (b) and (c) respectively. Note that  $A$ and $Y(z)$ are d-connected given $X$ in the twin-network  
by two different d-connecting paths,\footnote{Precisely: $A \leftarrow U_A \rightarrow A(z) \rightarrow X(z) \rightarrow Y(z)$ and  
$A \rightarrow X \leftarrow U_X \rightarrow X(z) \rightarrow  Y(z)$.}
However, in spite of this $A \,\ci\,Y(z)\mid X$ under the associated NPSEM-IE because
$X(z)=X$, and $A$ and $Y(z)$ are d-separated given $X(z)$ in the twin-network.
The SWIG ${\cal G}(z)$ shown in Figure  \ref{fig:second-torpedo}(c) makes manifest that $A$ is d-separated from $Y(z)$ given $X$,
hence $A \,\ci\,Y(z)\mid X$ under the FFRCISTG, hence also under the NPSEM-IE.
}

{
In addition, it may also be inferred from the SWIG that $A(z)  \,\ci\,Y(z)\mid X$;   
$A  \,\ci\,Y(z)\mid X(z)$ and $A(z)  \,\ci\,Y(z)\mid X(z)$ hold under the FFRCISTG (and hence also the NPSEM-IE). This is because it follows from causal irrelevance that, 
 given a SWIG ${\cal G}(a)$, if a label $a_i$ is present on some random node (equivalently if the SWIG contains a fixed node $a_i$), then $a_i$ may always be added to the label of any random node on which it is not already present. Consequently we are free to add $z$ to the label for $X$ and $A$ in ${\cal G}(z)$, from which these independences follow. 
 Note that in the Twin Network, although $A$ and $A(z)$ are d-separated from $Y(z)$ given $X(z)$, the path $A(z) \rightarrow X(z) \rightarrow Y(z)$ d-connects $A(z)$ and $Y(z)$ given $X$, hence we cannot read off
  $A(z)  \,\ci\,Y(z)\mid X$ from the Twin Network.
}

{
\citet{shpitser07hierarchy} provide an algorithm for merging nodes in a Twin Network, under a particular instantiation of the variables. This algorithm is conjectured to be complete for checking equality of the probability of counterfactual events.
A conditional independence statement corresponds to a (potentially exponential) set of equalities between probabilities of events. Thus, if the conjecture holds, then the algorithm of \citet{shpitser07hierarchy} provides a way to check counterfactual conditional independence implied by an NPSEM-IE. Though this approach is more involved, as noted earlier, in footnote \protect\ref{foot:twin}, 
it addresses a harder problem than SWIGs since it is determining all independencies implied by an NPSEM-IE model which also includes ``cross-world'' independencies. 
}

\subsection{Weaker causal models to which the po-calculus also applies}\label{sec:weaker}

In Section \ref{sec:po-calc}, we chose to express rule 3 of the po-calculus on the distribution level: $p(Y(x,z)) = p(Y(x))$, although the equality holds on the individual level: $Y(x,z) = Y(x)$ under the FFRCISTG, see also footnotes \ref{foot:weaker-swigs} and \ref{foot:r3}.
We chose to do so for several reasons.  First, this form is closer in spirit to Pearl's original formulation of the {\it do}-calculus.

Second, the weaker equality is expressible in the language of interventions, say via the do operator: $p(Y \mid \text{do}(x,z)) = p(Y \mid \text{do}(x))$.  This allows us to apply this rule, and other rules of po-calculus to causal models that are not counterfactual, but which allow discussion of interventional distributions, such as the \emph{agnostic causal model} of
\cite{spirtes01causation} which is \emph{defined} by the relationship between the observed data distribution and interventional distributions given by the extended g-formula (\ref{eqn:g}) re-expressed via the {\it do} operator.
Indeed, the FFRCISTG and the NPSEM-IE imply all distribution-level interventional statements that hold under the agnostic causal model, and these are the only statements that are relevant for the purposes of identification of interventional quantities expressible by the {\it do} operator.
Note that the distribution level equality has a graphical representation via {\it population SWIGs} in which missing edges correspond to the absence of population level direct effects, whereas the individual level counterfactuals are not necessarily the same.  See also \citep[Section 7]{thomas13swig}.

\subsection{Completeness Proofs}

Here we describe a number of completeness results referred to in the main body of the paper.  Before doing so, we state necessary preliminaries.
Given an acyclic directed mixed graph (ADMG) ${\cal G}(V)$ and a set $S \subseteq V$, an induced subgraph ${\cal G}(V)_S$ is defined to be a graph containing vertices $S$, and all edges in ${\cal G}(V)$ between elements in $S$.

Given an acyclic directed mixed graph (ADMG) ${\cal G}(V)$, we define a set $W \subseteq V$ to be fixable if $W = \emptyset$, or $W = \{ W_1, W_2, \ldots \}$ and there exists a set of ADMGs ${\cal G}_0(V)$, ${\cal G}_1(V \setminus \{ W_1 \})$, ${\cal G}_2(V \setminus \{ W_1, W_2 \})$, $\ldots$, ${\cal G}_k(V \setminus W)$, such that
\begin{itemize}
\item ${\cal G}_0(V) = {\cal G}(V)$.
\item For every $i = 0, \ldots, k-1$, $W_{i+1}$ has no element $V_j \in V \setminus \{ W_1, \ldots, W_i, W_{i+1} \}$ with a directed path from $W_{i+1}$ to $V_j$ and a path consisting exclusively of bidirected edges from $W_{i+1}$ to $V_j$ in ${\cal G}_i$.
\item For every $i = 1, \ldots, k$,  ${\cal G}_i(V \setminus \{ W_1, \ldots, W_i \})$ is obtained from ${\cal G}_{i-1}(V \setminus \{ W_1, \ldots, W_{i-1} \})$ by removing $W_i$ and all edges adjacent to $W_i$.
\end{itemize}
If $W \subseteq V$ is fixable, the set $S \equiv V \setminus W$ is said to be \emph{reachable}.  A set $S$ reachable in ${\cal G}(V)$ is said to be \emph{intrinsic} if the vertices in ${\cal G}(V)_S$ form a bidirected connected set.  Note the relationship between reachable sets and the precondition for Proposition \ref{prop:split}.

We have the following result.

\begin{thm}
Fix possibly intersecting sets $Y,A$ such that $Y(a)$ is ancestral in the SWIG ${\cal G}(V(a))$.  Then
\begin{align}
\notag
\MoveEqLeft{p(Y(a) = v_Y)} = \prod_{D \in {\cal D}(\GYa)} p\!\left(\!V_{D}(a,v_{\pas_{D}^{\GYa}}) = v_{D}\!\right),
\label{eqn:line-4}
\end{align}
and $p(Y(a))$ is not identified if there exists $D \in {\cal D}(\GYa)$ such that no inductive sequence of applications of Proposition \ref{prop:split} exists where every element $V_j \in A \cup (V \setminus D)$ is split such that the precondition of Proposition \ref{prop:split} is satisfied at every step, and $V_j(a)$ is marginalized from the resulting SWIG whenever $V_j \not\in D$.
\end{thm}
\begin{prf}
Assume such a set $D$ exists.  Assume $D$ is not a reachable set in ${\cal G}(V)$.  Then the results in \cite{richardson17nested} imply that there exists a hedge for $p(Y(a))$ and that $p(Y(a))$ is not identified \cite{shpitser06id}.

Assume $D$ is a reachable set, but some element $A_i \in D$ cannot be split by applying Proposition \ref{prop:split}.  This implies there exists a set of vertices
$W_1, \ldots, W_k$ in $D$ that are bidirected connected, and $W_k$ is a child of $A_i$ in ${\cal G}(V)$.  
Since $W_1, \ldots, W_k$, being elements of $D$, are in the set of ancestors of $Y$ in ${\cal G}(V(a))$, the sets $\{ A_i \}$, and
$\{ A_i, W_1, \ldots, W_k \}$ form a hedge for $p(Y(a))$, so $p(Y(a))$ is not identifiable.
\end{prf}

\begin{thm}
Fix subsets $Y,Z,A$ of $V$, in some ADMG ${\cal G}(V)$, where $Y,Z$ are disjoint, but may both intersect $A$.
Fix the largest subset $W \subseteq Z$, with $Z' = Z \setminus W$, such that $Z'(a,z')$ is m-separated from $Y(a,z')$ given $W(a,z')$ in ${\cal G}(V(a,z'))$, and let $A'$ be a maximal subset of $Z \cap A$ such that $A'(a,z')$ is m-separated from $Y(a,z')$ given $\{ W(a,z') : W \in Z \setminus (Z' \cup A') \}$.  Then $p(Y(a) \,|\, Z(a))$ is identified if $p(Y(a,z'), \{ W(a,z') : W \in Z \setminus (Z' \cup A') \})$ is identified.  If identification holds, we have:
\begin{align*}
p(Y(a) \,|\, Z(a)) =
\left.
\frac{
p(Y(a,z'), \{ W(a,z') : W \in Z \setminus (Z' \cup A') \})
}{
p(\{ W(a,z') : W \in Z \setminus (Z' \cup A') \})
} \right|_{Z \setminus A' = z_{Z \setminus A'}}.
\end{align*}
\end{thm}
\begin{prf}
If the stated assumptions hold, and $p(Y(a,z'), \{ W(a,z') : W \in Z \setminus (Z' \cup A') \})$ is identified, the conclusion follows by definition of conditioning.

Assume $p(Y(a,z'), \{ W(a,z') : W \in Z \setminus (Z' \cup A') \})$ is not identified.  It suffices to consider the case where
$p(\{ W(a,z') : W \in Z \setminus (Z' \cup A') \})$ is not identified.  The proof then follow the proof structure for analogous results in \cite{shpitser06idc,malinsky19po}, with the fact that $A \cap Y$ is potentially not an empty set not influencing the structure of the proof.

Non-identification of $p(\{ W(a,z') : W \in Z \setminus (Z' \cup A') \})$ implies the existence of a hedge, and the preconditions of the theorem imply the existence of an m-connecting (given $W$) path from an element in $W$ in the hedge to some element in $Y$.  Non-identification is established by induction on the structure of this path.  Specifically, fix an element $L$ on the path such that the inductive hypothesis that $p(L(a,z') | W'(a,z'))$ is not identified holds, where $W'$ is the subset of $W$ involved in the hedge, or in the m-connecting path from the hedge to $L$.  Thus, there exist two elements of the causal model that disagree on this distribution, but agree on the observed data distribution.  The induction then establishes that the distribution $p(L'(a,z') | W''(a,z'))$, where $L'$ is the next element on the m-connecting path, and $W''$ are all elements of $W$ that are either in the hedge, or witness m-connection of the path from the hedge to $L'$, is also not identified.  This is established by extending existing two elements with an appropriate distribution that yields a one to one mapping from distributions $p(L(a,z') | W'(a,z'))$ to distributions $p(L'(a,z') | W''(a,z'))$.
\end{prf}

\makeatletter\@input{xx_id.tex}\makeatother
\end{document}